\documentclass[a4paper,11pt]{article}
\usepackage{jheppubArXiv} 
\usepackage{lineno}

\usepackage{amsmath}
\usepackage{dsfont}
\usepackage{slashed}

\usepackage{physics}
\usepackage{multirow}
\usepackage{url}

\usepackage{graphicx}
\usepackage[table, svgnames, dvipsnames]{xcolor}
\usepackage{makecell, cellspace}

\usepackage{ragged2e}
\usepackage{array}
\usepackage{tabularx}
\usepackage{booktabs}

\definecolor{cset-aps-blueberry}{RGB}{28,128,158}
\definecolor{cset-aps-blue}{RGB}{46,44,184}
\definecolor{cset-aps-turquoise}{RGB}{0,67,88}
\definecolor{cset-aps-limegreen}{RGB}{190,219,67}
\definecolor{cset-aps-green}{RGB}{31,138,112}
\definecolor{cset-aps-yellow}{RGB}{255,225,25}
\definecolor{cset-aps-orange}{RGB}{253,116,0}
\definecolor{cset-aps-red}{RGB}{219,0,43}

\usepackage{hyperref}
\hypersetup{%
    colorlinks=true,
    linkcolor={cset-aps-red},
    linkbordercolor={cset-aps-red},
    filecolor={cset-aps-orange},
    filebordercolor={cset-aps-orange},
    citecolor={cset-aps-blue},
    citebordercolor={cset-aps-blue},
    urlcolor={cset-aps-green},
    urlbordercolor={cset-aps-green},
    menucolor={cset-aps-limegreen},
    menubordercolor={cset-aps-limegreen},
    breaklinks=true,
    pdfborderstyle={/S/U/W 2},
    pdfpagemode=UseOutlines,
    pdfstartpage={1},
}

\newcommand{\ii}{i}

\newcommand{\ie}{i.\,e., }
\newcommand{\eg}{e.\,g., }
\newcommand{\vect}[1]{\boldsymbol{#1}}

\newcommand{\cI}{c_{D^2}^{(f)}}
\newcommand{\cII}{c_{\mathcal{X}}^{(f)}}
\newcommand{\cIII}{c_{D^3}^{(f)}}
\newcommand{\cIV}{c_{D \mathcal{X}}^{(f)}}
\newcommand{\cV}{c_{D \tilde{\mathcal{X}}}^{(f)}}
\newcommand{\cVI}{c_{D^4}^{(f)}}
\newcommand{\cVII}{c_{\mathcal{X} \mathcal{Y}}^{(f)}}
\newcommand{\cVIII}{c_{D^2\mathcal{X}1}^{(f)}}
\newcommand{\cIX}{c_{\mathcal{X} \mathcal{\tilde{Y}}}^{(f)}}
\newcommand{\cX}{c_{D^2\mathcal{X}2}^{(f)}}
\newcommand{\cXI}{c_{D^2\mathcal{X}}^{(f)}}
\newcommand{\cXII}{c_{D^2\mathcal{X}3}^{(f)}}
\newcommand{\cXIII}{c_{\mathcal{X} \mathcal{Y} \sigma}^{(f)}}
\newcommand{\cXIV}{c_{\left\{ \mathcal{X} \mathcal{Y} \right\} }^{(f)}}
\newcommand{\cXV}{c_{\left\{ \mathcal{X} \mathcal{\tilde{Y}} \right\} }^{(f)}}

\newcommand{\gX}{g_{_{\mathcal{X}}}}
\newcommand{\gY}{g_{_\mathcal{Y}}}

\newcommand{\dia}{\mathfrak{E}}
\newcommand{\off}{\mathfrak{O}}

\newcommand{\orcid}[1]{\href{https://orcid.org/#1}{\includegraphics[width=7pt]{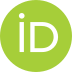}}}

\title{\boldmath Generalized Foldy-Wouthuysen approach for the derivation of non-relativistic effective field theories}

\author[a]{Tobias Asano\,\orcid{0000-0002-6257-8815},}
\author[b]{Fabio Di Pumpo\,\orcid{0000-0002-6304-6183},}
\author[c]{Enno Giese\,\orcid{0000-0002-1126-6352},}
\author[a,d]{and Motoaki Bamba\,\orcid{0000-0001-9811-0416}}
\affiliation[a]{Department of Physics, Graduate School of Engineering Science, Yokohama National University, 79-5 Tokiwadai, Hodogaya-ku, Yokohama 240-8501, Japan}
\affiliation[b]{Institut f{\"u}r Quantenphysik and Center for Integrated Quantum Science and Technology (IQST), Universit{\"a}t Ulm, Albert-Einstein-Allee 11, D-89081 Ulm, Germany}
\affiliation[c]{Technische Universit{\"a}t Darmstadt, Fachbereich Physik, Institut f{\"u}r Angewandte Physik, Schlossgartenstr. 7, D-64289 Darmstadt, Germany}
\affiliation[d]{Institute for Multidisciplinary Sciences, Yokohama National University, 79-5 Tokiwadai, Hodogaya-ku, Yokohama 240-8501, Japan}

\emailAdd{asano-tobias-gc@ynu.jp}
\emailAdd{tobias-asano@outlook.de}

\abstract{Effective field theories (EFTs) are a powerful framework for performing high-precision calculations at reduced complexity compared to their fundamental counterparts. 
A particularly important class of EFTs arises in the non-relativistic (NR) regime.
Their construction relies on a different realization of the underlying symmetries, since Lorentz invariance is no longer manifest in covariant form in the NR regime. 
This behavior imposes a link between certain matching coefficients, and therefore additional constraints, commonly referred to as hidden Lorentz invariance.
These constraints are established in quantum field theories on inertial flat spacetime, such as NR quantum electrodynamics.
However, deriving these constraints becomes considerably more involved for theories involving physics beyond the Standard Model or formulated in non-inertial spacetime backgrounds, where the hidden symmetry structure is less transparent. 
In this work, we present an approach to obtain the NR EFT by first constructing a relativistic EFT and then performing a generalized NR reduction based on an extended Foldy-Wouthuysen transformation. 
We illustrate this method by a quantum chromo-electrodynamics EFT for inertial flat spacetime, describing both electromagnetic and strong interactions, and show how it reduces to the established Lagrangian of NR quantum chromodynamics and electrodynamics. 
The hidden Lorentz invariance emerges as a direct consequence of the construction.
This approach provides a route to obtain the NR limits of more complex theories, \eg Dirac fields in non-inertial spacetime or extensions involving physics beyond the Standard Model. 
As an example, we apply the method to add the coupling of a pseudoscalar axion field in a simplified model and derive its NR limit.}

\begin{document}
\maketitle
\flushbottom

\section{Introduction}
While the Standard Model~\cite{Navas2024} provides the most precise experimentally verified description of particle physics to date, it may be viewed as an effective low-energy field theory valid up to some higher energy scale~\cite{Brivio2019}.
This viewpoint naturally motivates the construction of effective field theories~\cite{Weinberg1979,Georgi1993} (EFTs).
A prominent example in this context is the Standard Model Effective Field Theory (SMEFT)~\cite{Buchmueller1986,Grzadkowski2010,Lehman2014,Falkowski2023}, which parametrizes possible effects of physics beyond the Standard Model (BSM).
EFTs provide an effective description whenever one is interested only in a restricted set of observables or in phenomena within a specific energy regime, such that a description involving all degrees of freedom of the underlying theory is not required.
More generally, their construction relies on a hierarchy of physical scales, allowing for a systematic separation of the relevant low- and the irrelevant high-energy degrees of freedom.
Hence, the resulting EFTs reproduce predictions of the underlying theory for the observables within a specified energy regime, up to corrections organized in a systematic expansion in the ratio between the low- and high-energy scales.
The effects of high-energy degrees of freedom are then encoded in so-called \textit{Wilson coefficients}~\cite{Wilson1965,Wilson1974} multiplying operators in the effective Lagrangian.

A common realization of an EFT is obtained by integrating out heavy particles, thereby removing them from the set of dynamical degrees of freedom.
For example, in Low-Energy Effective Field Theory (LEFT)~\cite{Jenkins2018,Jenkins2018a,Liao2020,Murphy2021}, valid below the electroweak scale, the electroweak gauge bosons ($W^\pm$), ($Z$), and the Higgs field consequently no longer appear as dynamical degrees of freedom.
Formally, such a procedure corresponds to performing the functional integration over these degrees of freedom in the path integral, thereby generating an effective action for the remaining low-energy degrees of freedom~\cite{Wilson1974}.
More generally, the degrees of freedom that are integrated out need not necessarily be associated with heavy particles.
They may also correspond to high-energy momentum modes, referred to as \textit{hard modes}, while only the low-energy degrees of freedom relevant for the physical processes under consideration are retained in the EFT.
A particularly important class of examples is provided by systems in the non-relativistic (NR) regime, where characteristic particle momenta $\vect{p}$ are much smaller than the corresponding mass scale $m$, \ie  $|\vect{p}| \ll m$.
In such situations, NR EFTs~\cite{Caswell1986} such as non-relativistic quantum electrodynamics (NRQED)~\cite{Hill2013,Paz2015} and non-relativistic quantum chromodynamics (NRQCD)~\cite{Bodwin1995} provide systematic frameworks for precision calculations in atomic bound-states~\cite{Pineda1998a}, and heavy-quark systems~\cite{Lepage1992}.
In these theories, the effective expansion is organized in powers of the small non-relativistic parameter $v\sim \abs{\vect{p}}/m$, reflecting the hierarchy between the characteristic momentum and mass scales. In heavy-quark systems described by NRQCD, this hierarchy is typically accompanied by the additional dynamical scales $mv$ and $mv^2$.
For many NR applications, such EFTs provide some of the most precise descriptions available, combining a direct connection to the underlying fundamental theory with a significantly simpler theoretical framework.

Unlike relativistic EFTs such as SMEFT or LEFT, NR EFTs like NRQED and NRQCD require a separate treatment of particles and antiparticles.
They are typically constructed by writing down the most general effective Lagrangian from all operators consistent with the relevant symmetries (\ie all spacetime translation and rotationally invariant, gauge invariant, Hermitian, and parity ($P$), time-reversal ($T$), charge-conjugation ($C$) invariant operators), together with an appropriate and systematic power counting in the small expansion parameter.
The corresponding Wilson coefficients are subsequently determined through a matching to the underlying relativistic theory.
Although Lorentz invariance is no longer manifest in these NR theories, remnants of the underlying Poincar\'e symmetry survive and impose nontrivial constraints on the operator structure and on the Wilson coefficients.
In particular, boosts remain encoded in a nonlinearly realized form in the NR theory, implying relations among Wilson coefficients~\cite{Heinonen2012,Berwein2019}, often referred to as \textit{hidden Lorentz invariance}~\cite{Hill2013} or, in related contexts, reparametrization invariance~\cite{Luke1992}.
While these symmetry constraints are well understood in QED and QCD, their implementation becomes considerably more involved in theories with additional BSM fields or in non-inertial spacetime backgrounds, whether classical or quantum.
It is therefore advantageous to derive the NR limit directly from a relativistic EFT. 
Short-distance information already encoded in its Wilson coefficients can then be mapped systematically onto the NR theory, where hard contributions associated specifically with the scale separation between the relativistic and NR regimes must be included in the relativistic matching input if they are to be captured by this reduction without an additional matching step.
Such an EFT inherits the symmetries of the underlying relativistic theory, so that Lorentz symmetry remains manifest at the level of the starting Lagrangian.
As a consequence, no additional hidden-Lorentz constraints among otherwise independent Wilson coefficients need to be imposed at the relativistic level and Lorentz invariance is already built into the operator basis, thereby reducing the number of independent matching coefficients at a given order compared with a purely rotationally invariant NR parametrization.

\begin{figure}[ht]
    \centering
    \includegraphics[width=\textwidth]{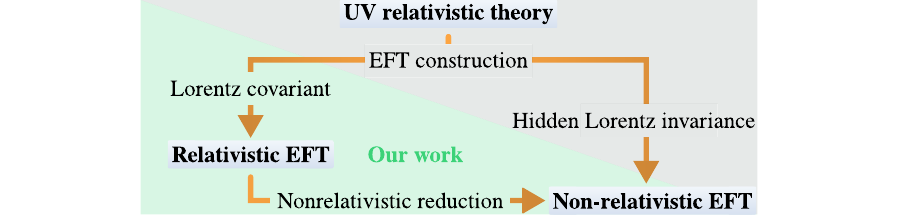}
    \caption{
    Schematic summary of the derivation of the NR limit of an EFT.
    As an alternative to constructing the theory directly at the NR level, where operators and fields are no longer manifestly Lorentz covariant (right-hand side), we first formulate an EFT at the relativistic, Lorentz-covariant level and relate it to the NR limit via an additional NR reduction based on an extended Foldy-Wouthuysen transformation. 
    While a direct NR construction implies relations among Wilson coefficients, also referred to as \textit{hidden Lorentz invariance}, starting from the relativistic theory generally involves fewer independent matching coefficients, and the corresponding relations emerge automatically from the NR reduction. 
    }
    \label{fig:1}
\end{figure}

In this paper, we explicitly demonstrate such a derivation as an alternative route to constructing NR EFTs such as NRQED and NRQCD in inertial flat spacetime.
The procedure is summarized in figure~\ref{fig:1}, and may in principle be extended to non-inertial spacetime or BSM physics.

First, we construct a relativistic EFT including both QCD and QED interactions, conceptually similar to LEFT but with a reduced set of matching coefficients, since we retain only operators consistent with the preserved discrete symmetries.
After reducing the operator basis of this relativistic EFT, we perform an NR reduction and show that it reproduces the NRQED and NRQCD Lagrangians.
NR reductions are an established technique, with the Foldy-Wouthuysen (FW) transformation~\cite{Foldy1950} (also known as Foldy-Wouthuysen-Tani transformation~\cite{Tani1951}) being a prominent and physically intuitive example.
This transformation is typically applied to Hamiltonians describing single-particle systems~\cite{Silenko2016}, but it is also used to derive operator structures in NR EFT Lagrangians~\cite{Gardestig2007}.
To the best of our knowledge, it has not yet been applied to an NR reduction starting with an EFT Lagrangian to reproduce NRQCD and NRQED.
We find that a naive implementation of the FW transformation in its most commonly used form~\cite{Foldy1950} fails to reproduce the correct matching structure, which necessitates a generalized scheme.

The benefit of starting with a relativistic Lagrangian is that the hidden Lorentz invariance emerges in a natural way.
In our framework, the NR reduction can be understood as a reorganization of the Wilson coefficients of the relativistic theory such that the known NR Wilson coefficients are reproduced.
For presentation purposes, we introduce our formalism with the help of the NR reduction of a specific Lagrangian, but we expect our approach to be applicable to any theory with a well-defined NR limit.
The paper is organized as follows:
In section~\ref{Sec:2}, we introduce the relativistic EFT describing QED and QCD interactions up to order $\Lambda^{-3}$ in the expansion in the characteristic energy scale $\Lambda$, which sets the regime of validity of the EFT, and present the reduced operator basis that serves as the starting point for the NR reduction.
The detailed construction of the complete relativistic operator basis and its subsequent reduction through Lorentz-covariant field redefinitions are given in appendix~\ref{App:A}.
In section~\ref{Sec:3}, we develop the generalized NR reduction scheme and apply it to the relativistic quantum chromo-electrodynamics (QCED) Lagrangian.
We first establish a consistent NR power counting and introduce the canonical, Foldy-Wouthuysen, and gauge-field redefinitions required for the reduction, before deriving the NR two- and four-fermion sectors.
In section~\ref{Sec:6}, we present two applications of the formalism: the first nontrivial EFT corrections to the NR Poincar\'e generators, and an extension of the framework to a simple BSM axion model, for which we derive the corresponding NR Lagrangian.
Finally, section~\ref{Sec:5} places our approach in the context of existing NR reduction methods, summarizes the main results, and discusses possible future applications.

\section{Relativistic effective field theory} \label{Sec:2}

As the derivation and construction of EFTs is an established technique~\cite{Weinberg1979,Georgi1993,Burgess2020}, we summarize only some key aspects in the following.
A common way to obtain an EFT is by integrating out degrees of freedom associated with scales above the regime of interest from an underlying theory, which need not be known explicitly. 
Formally, this procedure can be implemented in the path integral by integrating out heavy fields, and possibly also highly off-shell modes above the characteristic EFT scale $\Lambda$. 
In general, this approach produces a nonlocal effective action. Expanding this action in powers of external momenta and light masses divided by $\Lambda$ then yields a local effective Lagrangian valid at energies $E \ll \Lambda$.
In practice, the exact integration over heavy degrees of freedom can rarely be performed explicitly. 
Modern EFT constructions therefore proceed directly by writing down the most general local Lagrangian consistent with the symmetries of the underlying UV theory, organized order-by-order in the expansion parameter $E/ \Lambda$. 
For EFTs obtained by integrating-out heavy particles, the scale $\Lambda$ is typically associated with the mass scale of the lightest degrees of freedom that are no longer present as dynamical fields in the EFT.
The resulting EFT reproduces the low-energy predictions of the UV theory up to the desired order in the expansion parameter. 
For a specified set of low-energy degrees of freedom, symmetries, and power counting, including all local operators compatible with these assumptions at a given order yields the most general EFT description at that order. 
Their Wilson coefficients encode the UV physics and may receive both tree- and loop-level matching contributions.
The effective Lagrangian can therefore be written as $\mathcal{L}_\text{EFT} = \mathcal{L}_0 + \sum_{k=1}^N \Lambda^{-k} \sum_j c_{k,j} O_{k,j}$, where $\mathcal{L}_0$ denotes the leading-order Lagrangian, usually consisting of all relevant and marginal operators, while $O_{k,j}$ denote higher-dimensional operators allowed by the symmetries of the original theory. 
Since symmetry arguments alone do not determine the prefactors of these operators, each operator is accompanied by a numerical coefficient $c_{k,j}$, referred to as \textit{Wilson} coefficients.
The Wilson coefficients encode the short-distance physics above the EFT scale and are determined by matching the EFT to the underlying UV theory. 
In practice, this matching is performed by requiring that suitably chosen Green’s functions or $S$-matrix elements agree in the low-energy limit up to the relevant order in the EFT expansion. 
While EFTs are generally non-renormalizable in the traditional sense, they remain renormalizable order-by-order in the expansion parameter.
Ultraviolet divergences arising within the EFT are absorbed into redefinitions of the Wilson coefficients and operator basis, for example using dimensional regularization together with a minimal subtraction scheme.

The relativistic EFT in this work, used as an example to perform the NR reduction, is conceptually similar to LEFT~\cite{Jenkins2018,Jenkins2018a,Liao2020,Murphy2021}. 
Specifically, we work in the spontaneously broken phase of the Standard Model at energies below the electroweak scale.
As a consequence, the heavy Higgs, $(W^\pm)$, and ($Z$) bosons, as well as the top quark, have been integrated out and the remaining gauge symmetry is $SU(3)_\text{c} \times U(1)_{\text{em}}$ associated with QCD and QED interactions.
In addition, the EFT obeys the standard inertial flat-spacetime Poincar\'e symmetry, namely spacetime translations and Lorentz transformations.
For simplicity, we restrict ourselves to a flavour-diagonal situation and neglect mixing between fermion generations. 
Furthermore, we consider a subset of operators that conserve parity ($P$), time-reversal ($T$), and charge conjugation ($C$), \ie we set Wilson coefficients which would violate these conditions to zero.
The characteristic scale $\Lambda$ of this EFT is associated with the heavy degrees of freedom integrated out, and we assume $m_f \ll \Lambda$ for all fermions retained as dynamical fields.
Constructing an EFT is generally associated with renormalization and matching of Wilson coefficients.
In the present work, our main focus is the derivation of the non-relativistic limit of a relativistic Lagrangian.
Accordingly, renormalization-group evolution and matching are not explicitly addressed.
Furthermore, NR EFTs are usually constructed for specific processes with a small subset of particles and interactions, but to demonstrate the NR derivation in a general framework, we keep this aspect implicit, since we can always restrict the starting point to a subset of operators.

\subsection{Notation}
\label{subsec:Notation}

Before presenting the Lagrangian of our relativistic EFT, we introduce its building blocks and present the notation used in the paper.
Throughout the manuscript, we work in natural units $c=1=\hbar$ and in inertial flat-spacetime Minkowski metric $\eta^{\mu\nu}=(+1,-1,-1,-1)$ with mostly-minus sign convention.
The coupling of QED and QCD to left- and right-handed spinors is identical, which makes it most convenient to represent the Lagrangian with respect to Dirac four-spinors $f$ representing all fermion species, here the three generations of charged leptons $(e,\mu, \tau)$ and quarks $(u,c,d,s,b)$, together with their respective antiparticles, while the top quark is integrated out with the electroweak bosons.
Neutral Dirac fermions can be incorporated straightforwardly.
Standard-Model neutrinos, which appear as left-handed Weyl fields in LEFT, would require the corresponding extension of the spinor formulation. 
Fermionic operators are conveniently organized in terms of Dirac bilinears $\bar{f} \bullet f$, where $\bar{f} = f^\dagger \gamma^0$ is the standard Dirac adjoint spinor.
Lorentz invariance requires all Lorentz indices of the complete operator to be contracted to form a scalar.
Together with the Dirac four-spinors we use the standard basis of the Dirac algebra $\Gamma = \left\{ \mathds{1}, \gamma_5, \gamma^\mu, \gamma^\mu \gamma_5, \sigma^{\mu\nu} \right\}$, representing scalar, pseudoscalar, vector, axial vector, and tensor elements, in total sixteen independent $4\times 4$ matrices.
The $\gamma$ matrices obey the Clifford algebra $\left\{ \gamma^\mu , \gamma^\nu \right\} = 2 \eta^{ \mu \nu}$ and define the other matrices $\sigma^{\mu \nu} = \ii \left[ \gamma^\mu , \gamma^\nu \right]/2$ as well as $\gamma_5 = \ii \gamma^0 \gamma^1 \gamma^2 \gamma^3$.
Later, we will choose the Dirac basis in which $\gamma^0$ is diagonal with respect to upper and lower components of the four-spinor.

Gauge invariance requires derivatives acting on charged fields to appear in gauge-covariant combinations. 
We conventionally use $\ii D_\mu$, which is also convenient for constructing Hermitian operators.
The action on fermion fields is $D_\mu f = \partial_\mu f + \ii e Z_f A_\mu f+ \ii g_s \lambda^a A_\mu^a /2  f$ with gauge couplings
$e,g_s$, gauge group generators $Z_f,\lambda^a/2$, and four-vector potentials $A_\mu, A_\mu^a$ corresponding to the photon ($U(1)$) and the gluon ($SU(3)$), respectively.
Here, $Z_f$ denotes the charge number of the fermion and $\lambda^a$ are the eight Gell-Mann matrices labeled by $a$ running from $1$ to $8$. 

We choose for the paper a unified notation for the gauge fields by introducing variables $\mathcal{X},\mathcal{Y}$, which take the values $\mathcal{G}$ for the gluon, and $\mathcal{F}$ for the photon each. 
Within this notation, we abbreviate the two gauge couplings via $g_\mathcal{X}=(g_\mathcal{G}, g_\mathcal{F})=(g_s,e)$, the four-vector potential as $A_{\mathcal{X},\mu}^{a} = (A_{\mathcal{G},\mu}^{a}, A_{\mathcal{F},\mu}^{a})=(A_\mu^a, A_\mu)$, and the generators as $T_\mathcal{X}^a = (T_\mathcal{G}^a, T_\mathcal{F}^a)=(\lambda^a/2, Z_f)$.
The gluon generators satisfy the orthogonality relation $\Tr{T_\mathcal{G}^a T_\mathcal{G}^b}=\delta^{ab}/2$.
Note that we suppress the index $f$ of the generators $T_\mathcal{F}^a=Z_f$ to keep the notation compact.
Whenever we sum the four-vector potential with the corresponding group generators, we write script quantities, \ie $\mathcal{A}_{\mathcal{X},\mu}^a = T_\mathcal{X}^a A_{\mathcal{X},\mu}^a$. 
As a consequence, we can express the gauge covariant derivative on fermion fields as $D_\mu f = \left( \partial_\mu + \ii g_\mathcal{X} \mathcal{A}_{\mathcal{X},\mu}^a \right)f$ with the convention that whenever at least two quantities with the same script index $\mathcal{X}$ or $\mathcal{Y}$ appear, we sum over the two contributions $\mathcal{G}$ and $\mathcal{F}$.

Through the gauge-covariant derivative, we can define the field strength tensors of the gauge fields.
In our notation they take the compact form $\left[ D_\mu, D_\nu \right] f= \ii \gX \mathcal{X}_{\mu \nu} f$, where once again $\mathcal{X}_{\mu\nu} = (\mathcal{G}_{\mu \nu}, \mathcal{F}_{\mu \nu})$ contains both field strength tensors over which we sum due to the additional appearance of $\gX$.
Following these definitions, electric and magnetic fields associated with the gauge field $\mathcal{X}$ are $\mathcal{X}^{0i} f = - \mathcal{E}^i_\mathcal{X} f = - T_\mathcal{X}^{a} E_{\mathcal{X}}^{a,i}f $ and $\mathcal{X}^{ij} f = - \varepsilon^{ijk} \mathcal{B}^k_\mathcal{X} = - \varepsilon^{ijk} T_\mathcal{X}^{a} B_{\mathcal{X}}^{a,k}f $, with the totally antisymmetric Levi-Civita symbol in the $\varepsilon^{123}=+1$ convention.
In addition to $\mathcal{X}_{\mu\nu}$, we define its dual field strength tensor $\tilde{\mathcal{X}}^{\mu \nu} = \varepsilon^{\mu \nu \alpha \beta} \mathcal{X}_{\alpha \beta}/2$ with the same convention $\varepsilon^{0123}=+1$. 
Hence, electric and magnetic fields appear exchanged, \ie we have $\tilde{\mathcal{X}}^{0i} = - \mathcal{B}^{a,i}_\mathcal{X}$ and $\tilde{\mathcal{X}}^{ij} = + \varepsilon^{ijk} \mathcal{E}_\mathcal{X}^{a,k}$.

\subsection{EFT Lagrangian}

We organize the EFT Lagrangian into sectors according to the number of fermion fields they contain $\mathcal{L}_j = \bar{f}_1 f_1 ... \bar{f}_j f_j$ such that the full Lagrangian is expressed by $\mathcal{L} = \mathcal{L}_2 + \mathcal{L}_4 + \mathcal{L}_\text{Gauge}$, where $\mathcal{L}_\text{Gauge}$ contains only gauge fields.
Fermion sectors $\mathcal{L}_6$ and beyond exist in general, but within the relevant order of $\Lambda^{-3}$ no such EFT terms are allowed.
Finding the EFT Lagrangian for each sector requires all independently allowed local, Hermitian, Poincar\'e-, $P$-, $T$-, and $C$-symmetry preserving operators at each order.
In appendix~\ref{App:A}, we address this construction in more detail and present the Lagrangian up to $\Lambda^{-3}$ including the full set of operators, which we also reduce through Lorentz-covariant field redefinitions~\cite{Kamefuchi1961,Arzt1995} to a smaller basis.
Here, we present only this reduced quantum chromo-electrodynamics (QCED) Lagrangian
\begin{align} \label{Eq:LTinv}
    \mathcal{L}= \sum_f \bar{f} &\Bigg(i \slashed{D} - m_f - \gX \cII \frac{\mathcal{X}_{\mu \nu} \sigma^{\mu \nu}}{4\Lambda} + \gX \cIV  \frac{\left[ D^\mu, \mathcal{X}_{\mu \nu} \right] }{\Lambda^2} \gamma^\nu -\gX \cXI \frac{ \left\{ D_{\mu} , \left\{ D^\alpha , \mathcal{X}_{\alpha \nu} \right\} \right\}}{2 \Lambda^3} \sigma^{\mu \nu } \notag \\
    &+ \gX \gY \left[ \cVII \frac{\mathcal{X}^{\mu \nu} \mathcal{Y}_{\mu \nu}}{4\Lambda^3} + \ii  \cIX \frac{\mathcal{X}^{\mu \nu} \mathcal{\tilde{Y}}_{\mu \nu}}{4 \Lambda^3} \gamma_5 + \ii \cXIII \frac{  {\mathcal{X}_\mu}^{ \alpha} \mathcal{Y}_{\alpha \nu} }{2\Lambda^3} \sigma^{\mu \nu} \right] \notag \\
    & + \gX \gY \left[ \cXIV \frac{ \text{Tr} \left\{ \mathcal{X}^{\mu \nu} \mathcal{Y}_{\mu \nu} \right\}}{4\Lambda^3} - \ii  \cXV \frac{ \text{Tr} \left\{ \mathcal{X}^{\mu \nu} \mathcal{\tilde{Y}}_{\mu \nu} \right\} }{4 \Lambda^3} \gamma_5 \right] \Bigg)f \notag \\
    &  + \sum_{f,g}  \sum_{\Gamma} \frac{   d_{\Gamma,\mathcal{X}}^{(fg)} ( \bar{f} T^a_{\mathcal{X}} \Gamma f ) (\bar{g} T^a_\mathcal{X} \Gamma g ) }{\Lambda^2 } +  \sum_{f \neq g} \frac{ d_{5,\mathcal{X}}^{(fg)} \qty( \bar{f} T^a_\mathcal{X} \gamma_\nu f) D_\mu \qty( \bar{g} T^a_\mathcal{X} \sigma^{\mu\nu} g) }{\Lambda^3} \notag \\
    & -\frac{F^{\mu \nu} F_{\mu \nu}}{4} - \frac{\text{Tr} \left\{ \mathcal{G}^{\mu \nu} \mathcal{G}_{\mu \nu} \right\}}{2} +\ii g_s c_{\mathcal{G}^3} \frac{ \text{Tr} \left\{ {\mathcal{G}^{\mu}}_{\nu} {\mathcal{G}^{\nu}}_{\alpha} {\mathcal{G}^\alpha}_{\mu} \right\} }{\Lambda^2},
\end{align}
where the first two lines correspond to $\mathcal{L}_2$, the third line is $\mathcal{L}_4$ and the last line is $\mathcal{L}_\text{Gauge}$.
We use the Feynman slash notation $\slashed{D}= D_\mu \gamma^\mu$, $m_f$ is the fermionic mass, and the Wilson coefficients are labeled according to the operators they belong to.
To keep the notation compact, we sum over all four-fermion combinations in the first terms of the third line and define $d_{\gamma,\mathcal{X}}^{(ff)} = 0$ instead to account for the eliminated terms.\footnote{We keep the same notation for the fields, the masses,  and the Wilson coefficients, even though they differ from the ones in eq.~\eqref{Eq:1} introduced before the reduction to the reduced basis.}
The relativistic EFT considered here corresponds to a symmetry-restricted QCD+QED sector of LEFT, written in a Dirac-bilinear basis and further reduced by field redefinitions, where the latter is not unique.
Comparing with the corresponding LEFT basis, we find similar operator combinations and the number of independent operators are expected to the same.
Since this Lagrangian is still completely Lorentz covariant, a matching is straightforward and ideally performed at this stage with the reduced basis.
For terms proportional to one gauge coupling $\gX$, we chose the normalization such that a matching via a one-gauge-field process yields a direct correspondence to form factors, \ie one finds in the case of QED the anomalous magnetic moment $m_f c_{\mathcal{X}}^{(f)} / \Lambda = F_2^{(f)} \left(0 \right) = a^{(f)}_N$, a term proportional to the slope of the Dirac form factor $F_1$, and hence related to the electric charge radius $\frac{m_f^2}{\Lambda^2} \cIV - \frac{m_f^3}{\Lambda^3} \cXI= {F_1^\prime}^{(f)} \left( 0 \right)$ and $\frac{m_f^3}{\Lambda^3} \cXI = {F_2^\prime}^{(f)} \left( 0 \right)$.
Here, primes denote derivatives with respect to the dimensionless variable $q^2/m_f^2$, so that $F_i^\prime(0)$ is dimensionless.
The matching of Wilson coefficients associated with terms with two gauge fields and four fermion fields can be carried out by established methods~\cite{Pineda1998a,Hill2013}.

In the next section, we use this reduced Lagrangian as the starting point for non-Lorentz-covariant field redefinitions that yield the NR limit.
Although manifest Lorentz covariance of the fermion fields is lost, the full theory remains \textit{Lorentz invariant} by construction.

\section{Non-relativistic reduction} \label{Sec:3}
In this section, we derive the NR limit of the Lorentz-covariant EFT for the QCED Lagrangian by decoupling the upper and lower components of the Dirac spinor $\psi=(\phi,\chi)$ with respect to the Dirac basis, where $\gamma^0 = \begin{pmatrix}
    \mathds{1} & 0 \\ 0 & -\mathds{1}
\end{pmatrix}$,
$\gamma_5 = 
\begin{pmatrix}
    0 & \mathds{1} \\ 
    \mathds{1} & 0
\end{pmatrix}$, and 
$\gamma^i = 
\begin{pmatrix}
    0 & \sigma^i \\
    -\sigma^i & 0
\end{pmatrix}$.
Here, $\sigma^i$ are the Pauli matrices $\sigma^1 = \begin{pmatrix}
    0 & 1 \\ 1 & 0 \end{pmatrix}$, $\sigma^2 = \begin{pmatrix}
    0 & -\ii \\ \ii & 0 \end{pmatrix}$, and $\sigma^3 = \begin{pmatrix}
    1 & 0 \\ 0 & -1 \end{pmatrix}$.
When we separate the Lagrangian into even (diagonal) $\dia$ and odd (off-diagonal) $\off$ operators, the latter are responsible for the coupling between the components $\phi$ and $\chi$.\footnote{The matrices $\mathds{1}$, $\gamma^0$, and $\gamma^i \gamma_5$ are diagonal, while $\gamma_5$, $\gamma^i$, and $\gamma^0 \gamma_5$ are off-diagonal.}

While the field redefinitions that we will apply to diagonalize the Lagrangian, eq.~\eqref{Eq:LTinv}, correspond to a large extent to the established FW transformation~\cite{Foldy1950}, our approach shows that one cannot apply the FW transformation straightforwardly to all terms in the EFT Lagrangian in order to obtain a consistent NR limit.
To discuss necessary generalizations, we first introduce two types of leading-order manipulations on the Lagrangian and establish a consistent power-counting scheme for masses in the NR limit.
This procedure is closely related to how the FW transformation must be adjusted to introduce a generalized scheme.
After these generalizations, we return to eq.~\eqref{Eq:LTinv} and apply the reduction.

\subsection{Leading-order Lagrangian manipulations} \label{Sec:LOLag}
The diagonalization of the Lagrangian is implemented through field redefinitions $f=\exp{\lambda} f^\prime$ that have to be applied to the complete Lagrangian.
However, due to the order-by-order diagonalization, the actual diagonalization mechanism is entirely included in the modified NR leading-order (LO) fermionic Lagrangian
\begin{align}
\mathcal{L}_\text{NRLO} = \sum_f \bar{f} \left( \ii \gamma^0 D_0 - m_f \right) f = \sum_f \bar{f}^\prime \left( \ii \gamma^0 D_0 - m_f + \Delta_\text{time} + \Delta_\text{mass} \right) f^\prime
\end{align}
after the field redefinition, and the remainder of the Lagrangian only introduces higher orders.
The Lagrangian changes its form due to the time derivative and the mass term by $\Delta_\text{time} = \gamma^0 \left( \exp{ \lambda^\dagger} \ii D_0 \exp{\lambda} - \ii D_0 \right)$ and $\Delta_\text{mass} = - m_f \left( \exp{ \gamma^0 \lambda^\dagger \gamma^0 } \exp{ \lambda } - \mathds{1} \right)$, respectively.
Generally, we can introduce a field redefinition such that either $\Delta_\text{time}$ or $\Delta_\text{mass}$ eliminates a term in the EFT Lagrangian to derive the NR limit.
As an example, consider the term $\bar{f} \gamma^0 \left[D_0, \vect{\mathcal{E}}_\mathcal{X} \cdot \vect{\alpha} \right] / m_f^2 f$. Generally, we can either choose $\lambda =\ii \vect{\mathcal{E}}_\mathcal{X} \cdot \vect{\alpha} / m_f^2$, such that $\Delta_\text{time}$ eliminates the term in exchange for new contributions from $\Delta_\text{mass}$, as well.
Conversely, if choose $\lambda = \gamma^0 \left[ D_0, \vect{\mathcal{E}}_\mathcal{X} \cdot \vect{\alpha} \right] / \big( 2 m_f^3 \big)$ then $\Delta_\text{mass}$ is responsible for the elimination and we obtain new contributions from $\Delta_\text{time}$.

\subsection{Non-relativistic counting rule} \label{Sec:CR}

To diagonalize the Lagrangian $\mathcal{L}_2= \sum_f \bar{f} \left( \ii \gamma^0 D_0 - m_f + \dia + \off \right) f$, we separate its contributions into even and odd operators.\footnote{A similar separation and discussion can also be made for $\mathcal{L}_4$, which we keep implicit in the following.}
The FW field redefinition $f=\exp{\off /\left(2 m_f \right)} f^\prime$ uses the fact that $\Delta_\text{mass} = -m_f \left( \exp{\off /m_f}-\mathds{1}\right)$ and therefore the Lagrangian given by $\mathcal{L} = \sum_{f} \bar{f}^\prime \left( \ii \gamma^0 D_0 - m_f + \dia^\prime + \off^\prime / m_f \right) f^\prime$ has a new odd part which is now suppressed by one order further in $m_f$.
This procedure can be repeated until the desired order in $m_f$ is reached.

So far, the expansion of the Lagrangian is solely determined by inverse powers of $\Lambda$.
However, since the core of the FW field redefinition relies on generators $\lambda$ as a function of the fermionic rest mass $m_f$, the Lagrangian becomes reorganized with respect to the new natural NR scale $m_f$ instead of $\Lambda$.
Furthermore, the exponential in the FW field redefinition $f=\exp{\off /\left(2 m_f \right)} f^\prime$ defines only a perturbative series, and is therefore only meaningful if $\off/m_f$ is small with respect to matrix elements of external fields.\footnote{This condition implicitly restricts states of fermions and gauge fields to NR momenta, \eg $\abs{\vect{p}_f} \ll m_f$.}
As a result, we count the NR Lagrangian in orders of $\Lambda^{-r} m_f^{-u}$ with $r,u$ integers.
We truncate the relativistic EFT independently at $\mathcal{O}(\Lambda^{-3})$ and the subsequent NR expansion at $\mathcal{O}(m_f^{-3})$. 
As a result, all operators will be consistent up to $m_f^{-3}$ and the NR Wilson coefficients will be corrected up to $\Lambda^{-3}$.
This counting is only consistent if the generators of the field redefinitions $\lambda$ and the terms in the Lagrangian do not contain implicit mass factors.
However, every appearance of time derivatives within EFT terms, at least first-order ones on fermion fields and at least second-order ones on gauge fields, spoils this counting rule.
In the following, we explain this issue and its resolution.

\subsubsection{EFT terms with time derivatives on fermion fields}
\label{Sec:fermion-derivative}
Any EFT term with at least a single time derivative acting on a fermion field can be expressed as $\bar{f} \left\{ \ii D_0, \off + \dia \right\} f$ with unspecified, perturbative $\off$ and $\dia$.\footnote{If an EFT term with a time derivative acting on the fermion field is not already in an anticommutator form, it can always be cast into it; the difference is then only proportional to commutators between $D_0$ and the fields contained in $\dia$ and $\off$.}
In this anticommutator form, $\dia$ and $\off$ can still contain additional time derivatives on the fermion field.
Using then an FW redefinition of $f$ with a generator $\lambda=  \left\{ \ii D_0, \off \right\} / (2m_f)$ eliminates the odd part in the Lagrangian via $\Delta_\text{mass}$.
In this case, the time derivative is on-shell proportional to a mass factor through the EOM $\ii D_0 f = \left( m_f \gamma^0 + ... \right)f$, \ie the generator $\lambda = \left\{ \ii D_0, \off \right\} / (2m_f)$ is not of order $\off/m_f$, but rather of order $\off$, which is in conflict with the perturbative idea of the FW transformation.

Instead of the FW field redefinition, we can redefine the fermion field by a Hermitian generator $\lambda=-\gamma^0 \left( \off + \dia \right)$ leading to a non-unitary transformation.
In this case, $\Delta_\text{time}$ is responsible for both the even and odd part $\bar{f} \left\{ \ii D_0, \off + \dia \right\} f$ being eliminated in the Lagrangian. 
The remaining contributions from the field redefinition correspond exactly to the naive application of the EOM on the fermion field in the Lagrangian with $\ii D_0 f = \left( m_f \gamma^0 + ... \right)f$.
For higher-order time-derivative terms, we can repeat the procedure until all time derivatives are eliminated up to a desired order.
We call this type of field redefinition \textit{canonical reduction}, since the canonical conjugate momentum $\pi_f^\prime$ takes the standard form $\pi_f^\prime = \ii {f^\prime}^\dagger$.
Obtaining this conjugate momentum immediately implies that the redefined Lagrangian possesses an inner product that relates to the standard Born rule~\cite{Born1926,Mackey1957}, where fermionic excitations are counted according to $\int \dd[3]{x} \psi^\dagger \psi$.
Summarizing the above discussion, every time derivative acting on a fermion field in the Lagrangian counts as a mass factor, \ie an expression $\left\{ \ii D_0, \off \right\}/m_f$ of the order $\off$, and requires a canonical reduction.

\subsubsection{Second-order time derivatives and higher on gauge fields}

\begin{figure}[ht]
    \centering
    \includegraphics[width=\linewidth]{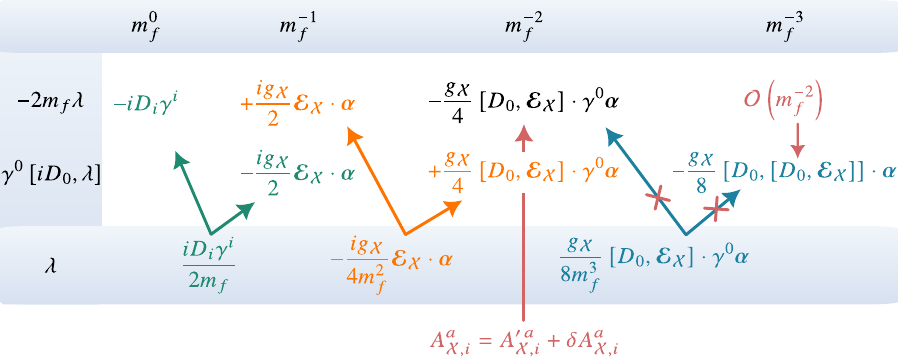}
    \caption{Traditional FW transformation of the term $\bar{f} \ii D_i \gamma^i f$. The first and second row show the change in the NRLO Lagrangian to lowest order, namely $-2m_f \lambda$ and $\gamma^0 \left[ \ii D_0, \lambda \right]$, due to the redefinition specified by $\lambda$ (third row).
    These new contributions are categorized in orders of $m_f^{-1}$ (columns). 
    Eliminating the initial $\ii D_i \gamma^i$ and subsequent terms (green and orange) eventually leads to $\left[ D_0, \vect{\mathcal{E}}_\mathcal{X} \right]$ appearing in $m_f^{-2}$. 
    If we continue with a standard FW redefinition (blue), then the new term in $m_f^{-3}$ is still $m_f^{-2}$, so we would introduce an inconsistency in our NR counting. Therefore, we are forced to redefine the gauge field (red).}
    \label{Fig:FWEfield}
\end{figure}

The second type of terms that cannot be diagonalized via the standard FW transformation are second- and higher-order time derivatives acting on gauge fields, \eg a term of the form $ \bar{f} \left\{ \left[D_0 , \mathcal{E}_\mathcal{X}^i \right], \off^i \right\} f$ with an arbitrary off-diagonal perturbative operator $\off^i$.
Figure~\ref{Fig:FWEfield} visualizes such a case with the example of an FW transformation of the odd term $\bar{f} \ii D_i \gamma^i f$ by standard LO Lagrangian manipulations and why it causes inconsistencies.

For simplicity we focus only on the effect of the FW field redefinition on the NRLO Lagrangian and restrict ourselves to the first term in the exponential series, \ie we consider $\Delta_\text{mass} = -2m_f \lambda$ and $\Delta_\text{time} = \gamma^0 \left[ \ii D_0, \lambda \right]$.
The generator $\lambda=\ii D_i \gamma^i / (2m_f)$ of the first field redefinition introduces $\Delta_\text{mass}=-\ii D_i \gamma^i$, eliminating the initial $+\ii D_i \gamma^i$ in the Lagrangian, while $\Delta_\text{time}$ is proportional to the electric field at order $m_f^{-1}$ (green part).
Subsequently, the electric-field term is eliminated by an analogous field redefinition giving rise to a time derivative on the electric field in $m_f^{-2}$ (orange part).
After another field redefinition, we end up with $\left[D_0,\left[ D_0, \mathcal{E}_\mathcal{X}^i \right] \right]$ in $m_f^{-3}$ (blue part). 
On-shell, however, $ \left[D_0,\left[ D_0, \mathcal{E}_\mathcal{X}^i \right] \right] \propto \sum_f \bar{f} m_f \gamma^0 \gamma^i f$, \ie even though this term seems to be of the order $m_f^{-3}$, it is actually $\mathcal{O}\left( m_f^{-2} \right)$ because the fermion EOM connects the time derivative with the fermion's mass.
Therefore, the diagonalization of the Lagrangian is not consistent within this order when using such a field redefinition.

This issue can be circumvented by performing a \textit{gauge-field redefinition} $A_{\mathcal{X},i}^a = A^{\prime \, a}_{\mathcal{X},i} +\delta A^a_{\mathcal{X},i}$ (red part) of the spatial components instead.
As intended, the gauge-field Lagrangian changes as
\begin{align} \label{Eq:Gauge}
    - \frac{X_{\mu \nu}^a X^{a, \mu \nu}}{4} = -  \frac{{X^\prime}^a_{\mu \nu} {X^\prime}^{a, \mu \nu}}{4} + \left[ D^\mu, {X^{\prime}}^a_{\mu i} \right] \delta A^{a,i}_\mathcal{X} + \frac{ \left( \left[ D_0 \vect{\delta A}_\mathcal{X} \right] \right)^2 - \left(  \left[ \vect{D} \times \vect{\delta A}_\mathcal{X} \right] \right)^2}{2} + \Delta \mathcal{L}_{\mathrm{NA}}^{(\geq 2)}
\end{align}
and the field redefinition applies the EOM in the EFT Lagrangian~\cite{Knetter1994}.
For a non-Abelian gauge field, additional terms starting at quadratic order in $\delta A$ are collected in $\Delta \mathcal{L}_{\mathrm{NA}}^{(\geq 2)}$. 
They do not affect the NR order considered here and are therefore omitted.
Due to time derivatives acting on the fermion fields in eq.~\eqref{Eq:Gauge}, we have to redefine the fermion fields again to keep the NR counting consistent, see the discussion in the previous subsection~\ref{Sec:fermion-derivative}.

On the other hand, the opposite case of a diagonal EFT term $ \bar{f} \left[D_0 , \mathcal{E}_\mathcal{X}^i \right] \dia^i f$ can be eliminated through a diagonal unitary field redefinition.
This procedure works because $\Delta_\text{mass} = 0$ vanishes for this field redefinition and only $\Delta_\text{time}$ eliminates diagonal terms.

As a general rule for a consistent NR diagonalization, first-order time derivatives of fermions and second-order ones of gauge fields must not be eliminated through an FW field redefinition.
Instead, the field redefinition must apply their corresponding EOMs.

\subsection{Generalized NR reduction scheme} \label{Sec:Gen}

Motivated by the discussion from the previous subsection, we present the general NR reduction scheme in figure~\ref{fig:FWroadmap} that yields a consistent NR counting up to the desired order:
\begin{figure}[ht]
    \centering
    \includegraphics[width=\textwidth]{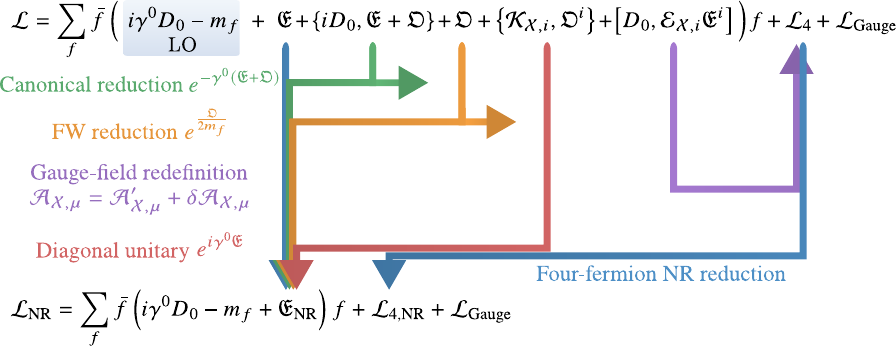}
    \caption{Generalized NR reduction scheme. Each fermion sector is diagonalized in increasing order, starting with the two-fermion sector. First, the canonical reduction (green) removes time derivatives acting on fermion fields. Standard FW field redefinitions (orange) then eliminate odd operators. Next, a gauge-field redefinition replaces odd EFT terms proportional to the pure gauge-field EOM, \ie on-shell $\mathcal{K}_{\mathcal{X},i} =j_i$, by fermionic currents $j_i$. This way, EFT terms with second-order time derivatives on gauge fields $\left[ D_0, \mathcal{E}_{\mathcal{X}_i}\right]$ are shifted into higher fermion sectors (purple).
    Finally, a diagonal fermion-field redefinition (red) removes diagonal terms of the form $\left[ D_0, \mathcal{E}_{\mathcal{X}_i} \mathfrak{E}^i \right]$. The same procedure can be applied analogously to the other fermion sectors beginning with $\mathcal{L}_4$. }
    \label{fig:FWroadmap}
\end{figure}
Since we organize the NR reduction in a way that only sectors with more fermion fields are affected but never sectors with less fermion fields, we naturally begin with $\mathcal{L}_2$ and continue in increasing order.
Regarding the two-fermion sector, it can always be organized according to figure~\ref{fig:FWroadmap}, where the NRLO Lagrangian is followed by a diagonal and off-diagonal part $\dia$ and $\off$, respectively, containing no time derivatives on fermion fields and only first-order time derivatives on gauge fields.
Any term with at least one time derivative on fermion fields can be cast into the form $\left\{ \ii D_0, \dia + \off \right\}$.
The remaining two terms are the off-diagonal and diagonal EFT terms with at least second-order time derivatives on gauge fields.
It is advantageous to rewrite any EFT term featuring $\left[ D_0, \mathcal{E}_{\mathcal{X},i} \right]$ into the form $\mathcal{K}_{\mathcal{X},i}$ representing the pure gauge-field EOM, such that on-shell $\mathcal{K}_{\mathcal{X},i} = j_i$ is only proportional to currents $j_i$ composed of fermion fields.
For QED, \eg we have $\mathcal{K}_{\mathcal{F},i} = \partial^\mu \mathcal{F}_{\mu i}$.
As such, a gauge-field redefinition applying the EOM modifies only sectors with more fermion fields.

From this starting point we first perform the canonical reduction (green part in figure~\ref{fig:FWroadmap}) with a Hermitian generator $\lambda = - \gamma^0 \left( \dia + \off \right)$ and repeat this procedure until no terms with time derivatives acting on fermion fields remain at the order of interest to ensure a consistent NR counting.
Next, we eliminate the odd part $\off$ via standard FW redefinitions (orange part in figure~\ref{fig:FWroadmap}) in such a way that only terms with time derivatives on gauge fields remain as prescribed in figure~\ref{fig:FWroadmap}.
Applying a gauge-field redefinition (purple part in figure~\ref{fig:FWroadmap}) shifts the two-fermion term entirely into higher fermion sectors beginning with $\mathcal{L}_4$.
Finally, the remaining diagonal EFT terms with second-order time derivatives on gauge fields are redefined by a diagonal unitary field redefinition (red part in figure~\ref{fig:FWroadmap}), such that the two-fermion sector is diagonalized up to the desired order $\mathcal{L}_\text{2,NR} = \sum_f \bar{f} \left( \ii \gamma^0 D_0 - m_f + \dia_\text{NR} \right) f$ and the NR limit follows.
The four-fermion sector and beyond can be diagonalized using the same procedure but then the corresponding field redefinitions of fermion fields depend on multiple fermion fields, for example $f = \exp{ \lambda \left( \bar{g} g \right)} f^\prime$ for the case of four fermion fields.
In section~\ref{Sec:4ferm} we discuss the diagonalization of four-fermion terms in more detail when applying this scheme to our relativistic Lagrangian. 

\subsection{NR reduction of the \texorpdfstring{$\Lambda^{-3}$}{Λ⁻³} QCED Lagrangian} 

In this section, we diagonalize the Lagrangian from eq.~\eqref{Eq:LTinv} and derive its NR limit within the generalized scheme developed in section~\ref{Sec:Gen}.

\subsubsection{Canonical reduction} \label{Sec:Canonical_red}

First, we perform the canonical reduction (green part in figure~\ref{fig:FWroadmap}) and eliminate the time derivatives on fermion fields contained in the noncanonical (NC) part
\begin{align} \label{Eq:HD}
    \mathcal{L}_{\text{NC}} =& - \sum_f \bar{f}  \gX \cXI \frac{\left\{ D_\mu, \left\{ D^\alpha, \mathcal{X}_{\alpha \nu} \right\} \right\} \sigma^{\mu \nu}}{2 \Lambda^3} f +  \sum_{f \neq g} \frac{ d_{5,\mathcal{X}}^{(fg)} \qty( \bar{f} T^a_\mathcal{X} \gamma_\nu f) D_\mu \qty( \bar{g} T^a_\mathcal{X} \sigma^{\mu\nu} g ) }{\Lambda^3}
\end{align}
of the Lagrangian from eq.~\eqref{Eq:LTinv}.
The relevant time derivatives in eq.~\eqref{Eq:HD} are given by $\mu=0$ or $\alpha=0$, which are eliminated by applying the field redefinition
\begin{align} \label{Eq:fermionFR}
\begin{split}
    f=&  \exp{-\gX \cXI \gamma^0 \frac{ \left\{ \ii D_0, \vect{\mathcal{E}}_\mathcal{X} \cdot \ii \vect{\alpha} \right\} - \left\{ \ii \vect{D} \cdot, \vect{\mathcal{B}}_\mathcal{X} \times \ii \vect{\alpha} \right\} - \left[ \vect{D} \cdot, \vect{\mathcal{E}}_\mathcal{X} \right]}{2\Lambda^3}} \\
    &\times \exp{-\sum_{g \neq f}  \frac{d_{5,\mathcal{X}}^{(gf)}T^a_\mathcal{X}  \gamma^i \left( \bar{g}^\prime T^a_\mathcal{X} \gamma^i g^\prime \right)}{\Lambda^3} }.
\end{split}
\end{align}
In eq.~\eqref{Eq:fermionFR}, we introduce the abbreviations $\left[ \vect{D} \cdot, \vect{\mathcal{E}}_\mathcal{X} \right]  = \left[ D_i , \mathcal{E}_\mathcal{X}^i \right] $,
an analogous definition for the anticommutator, and $\alpha^i = \gamma^0 \gamma^i$ together with the corresponding vector $\vect{\alpha} = ( \alpha^1, \alpha^2, \alpha^3)^\text{T}$.
The linear contribution of the exponentials in the first line of eq.~\eqref{Eq:fermionFR} eliminates all EFT terms with time derivatives acting on fermion fields in the Lagrangian. 
Redefining four-fermion terms in the Lagrangian requires a redefinition depending again on fermion fields like the second exponential in the first line of eq.~\eqref{Eq:fermionFR}, where primed fields denote the redefined fields.
After these redefinitions, the Lagrangian takes the form
\begin{subequations} \label{Eq:can}
\begin{align}
    \mathcal{L}_2= \sum_f \bar{f} &\Bigg(i \slashed{D} - m_f +\gX \left[ -\cII \frac{\mathcal{X_{\mu\nu} \sigma^{\mu\nu}}}{4\Lambda}  + \bar{c}_{D\mathcal{X}}^{(f)} \gamma^0 \frac{\left[ \vect{D} \cdot , \vect{\mathcal{E}}_\mathcal{X} \right] }{\Lambda^2} + \cIV \frac{ \left[ D^\mu , X_{\mu i} \right] \gamma^i}{\Lambda^2} \right] \notag \\
    &+ \gX \gY \left[ \bar{c}_{\mathcal{XY}}^{(f)} \frac{ \mathcal{X}_{\mu\nu} \mathcal{Y}^{\mu \nu} }{4\Lambda^3} + \bar{c}_{\mathcal{E}_\mathcal{X}\mathcal{E}_\mathcal{Y}}^{(f)} \frac{ \vect{\mathcal{E}}_\mathcal{X} \cdot \vect{\mathcal{E}}_\mathcal{Y}}{\Lambda^3} + \ii \cXIII \frac{  {\mathcal{X}_\mu}^{\alpha} \mathcal{Y}_{\alpha \nu} }{2\Lambda^3} \sigma^{\mu \nu} + \ii \cIX \frac{\mathcal{X}^{\mu \nu} \mathcal{\tilde{Y}}_{\mu \nu}}{4 \Lambda^3} \gamma_5 \right] \notag \\
    &- \gX \cXI  \left[ \frac{ \left\{ \ii \vect{D} \cdot , \left\{ \vect{D} \times, \vect{\mathcal{B}}_\mathcal{X} \right\} \right\} }{2\Lambda^3}  -\ii  \frac{\left\{ \left\{ \ii \vect{D} \cdot , \vect{\mathcal{E}}_\mathcal{X} \right\} , \ii \vect{D} \cdot \vect{\alpha}  \right\} - \left[ \left[ \vect{D} \cdot, \vect{\mathcal{E}}_\mathcal{X} \right], \vect{D} \cdot \vect{\alpha} \right] }{2\Lambda^3} \right] \notag \\
    &+ \gX \gY \frac{  \cXIV \text{Tr} \left\{ \mathcal{X}^{\mu \nu} \mathcal{Y}_{\mu \nu} \right\} - \ii \cXV  \text{Tr} \left\{ \mathcal{X}^{\mu \nu} \mathcal{\tilde{Y}}_{\mu \nu} \right\}  }{4\Lambda^3}   \Bigg) f 
\end{align}
where we suppress the primes on the fields for notational clarity.
New canonical Wilson coefficients $ \bar{c}_{D\mathcal{X}}^{(f)} = \cIV - \frac{m_f}{\Lambda} \cXI $, $\bar{c}_{\mathcal{E}_\mathcal{X}\mathcal{E}_\mathcal{Y}}^{(f)}=\frac{\cXI + c_{D^2\mathcal{Y}}^{(f)}}{2}$, and $\bar{c}_{\mathcal{XY}}^{(f)}  = \cVII +  4 \frac{\cXI+ c_{D^2 \mathcal{Y}}^{(f)}}{2}$ arise due to the redefinition.
In particular, the canonical Wilson coefficient $\bar{c}_{D\mathcal{X}}^{(f)}$ has a physical relevance because it is proportional to the form factor $F_1^\prime(0)$.
In addition, combining the cross product with an anticommutator $\left\{ \vect{D} \times ,\vect{\mathcal{B}}_\mathcal{X} \right\} = \varepsilon^{ijk} \left[ D_i, \mathcal{B}_\mathcal{X}^j \right] \vect{e}_k$ with Cartesian unit vector $\vect{e}_k$ reduces to a commutator between the respective components.

Applying the field redefinitions also to the four-fermion sector yields its canonical form
\begin{align}
\begin{split} \label{Eq:scattering3}
    \mathcal{L}_4 =& \sum_{f,g}  \sum_{\Gamma} \frac{   d_{\Gamma,\mathcal{X}}^{(fg)} ( \bar{f} T^a_\mathcal{X} \Gamma f ) (\bar{g} T^a_{\mathcal{X}} \Gamma g ) }{\Lambda^2 } + \sum_{f \neq g} \frac{m_f + m_g}{\Lambda} d_{5,\mathcal{X}}^{(fg)}  \frac{  ( \bar{f} T^a_\mathcal{X} \gamma^i f ) (\bar{g} T^a_{\mathcal{X}} \gamma^i g ) }{\Lambda^2 }  \\
    &+  \sum_{f \neq g} \Bigg\{ \frac{ d_{5,\mathcal{X}}^{(fg)} \qty( \bar{f} T^a_\mathcal{X} \gamma_\nu f) D_i \qty( \bar{g} T^a_\mathcal{X} \sigma^{i \nu} g) }{\Lambda^3} + \frac{ d_{5,\mathcal{X}}^{(fg)} \qty( \bar{f} T^a_\mathcal{X} \gamma^i f) \qty( \bar{g} T^a_\mathcal{X} \ii D_i g + \text{H.c.}) }{\Lambda^3} \\
    &\hspace{1.5cm}- \frac{ d_{5,\mathcal{X}}^{(fg)} \qty( \bar{f} T^a_\mathcal{X} \gamma^0 \vect{\alpha} f) \cdot  \vect{D} \times \qty( \bar{g} T^a_\mathcal{X} \vect{\Sigma} g)}{\Lambda^3} \Bigg\}.
\end{split}
\end{align}
\end{subequations}
Naturally, the gauge sector does not change under these field redefinitions.
As a consequence of the canonical reduction, the canonical Lagrangian does not contain additional time derivatives acting on fermion fields besides the NRLO time derivative, such that the conjugate momentum becomes the trivial $\pi_f = \partial \mathcal{L}/ \partial \left( \partial_t f \right) =  \ii f^\dagger$.
Enforcing the canonical form comes at the price of giving up Lorentz covariance with respect to the transformation behavior of fermion fields $f$.
Consequently, the individual NR operators are no longer manifestly Lorentz covariant, while Lorentz invariance of the full theory is preserved by the invertible field redefinitions.

\subsubsection{Non-relativistic two-fermion sector}

Next, we move to the orange part in figure~\ref{fig:FWroadmap} and apply an FW field redefinition
\begin{align} \label{Eq:Reduction}
\begin{split}
    f =& \exp{ \frac{\gamma_0}{2} \qty( \frac{ \ii \vect{D} \cdot \vect{\alpha}}{m_f} - \frac{ \qty( \ii \vect{D} \cdot \vect{\alpha})^3}{3 m_f^3})} \exp{-\ii \gY \qty(1+\frac{m_f}{\Lambda} c_\mathcal{Y}^{(f)}) \frac{ \vect{\mathcal{E}}_\mathcal{Y} \cdot \vect{\alpha}}{4m_f^2}} \\
    &\times \exp{ \gY \gamma_0 \frac{ \left\{ \vect{D} \times, \vect{\mathcal{B}}_{\mathcal{Y}} \right\} \cdot \vect{\alpha} + c_\mathcal{Y}^{(f)} \left\{ \ii \vect{D} \cdot, \vect{\mathcal{B}}_\mathcal{Y} \right\} \gamma_5 }{8m_f^3} } f^\prime.   
\end{split}
\end{align}
Note that we do not display field redefinitions at the order $m_f^{-4}$ that are required to eliminate odd operators at the order $m_f^{-3}$, since their elimination does not further modify the diagonal part of the Lagrangian.
The two-fermion Lagrangian reduces to $\mathcal{L}_2 = \sum_f \bar{f}^\prime \mathcal{I}_{f} f^\prime$ with
\begin{align} \label{Eq:L2}
    \mathcal{I}_{f} =&i \gamma^0 D_0 - m_f + \frac{\vect{D}^2}{2m_f} + \frac{\vect{D}^4 }{8m_f^3} + \gX \left[ \frac{ \left\{ \vect{D}^2, \vect{\mathcal{B}}_\mathcal{X} \cdot \vect{\Sigma} \right\} }{8m_f^3} + c_{\text{F},\mathcal{X}}^{(f)} \frac{\vect{\mathcal{B}}_\mathcal{X} \cdot \vect{\Sigma}}{2m_f} +c_{\text{D}, \mathcal{X}}^{(f)} \gamma^0 \frac{ \left[ \vect{D} \cdot, \vect{\mathcal{E}}_\mathcal{X} \right] }{8m_f^2} \right] \notag \\
    &+ \gX \left[ c_{\text{S},\mathcal{X}}^{(f)} \gamma^0 \frac{\left[ \ii \vect{D} \times, \vect{\mathcal{E}}_\mathcal{X} \right]\cdot \vect{\Sigma}}{8m_f^2} + c_{p^\prime p, \mathcal{X}}^{(f)} \frac{ \left\{ \vect{D} \cdot \vect{\Sigma}, \vect{\mathcal{B}}_\mathcal{X} \cdot \vect{D} \right\} }{8m_f^3} + c_{\text{M},\mathcal{X}}^{(f)} \frac{ \left\{ \ii \vect{D} \cdot, \left\{ \vect{D} \times, \vect{\mathcal{B}}_\mathcal{X} \right\} \right\} }{8m_f^3}  \right]  \notag \\
    &+ \gX c_{\text{W},\mathcal{X}}^{(f)} \frac{ \left[ D_i , \left[ D_i, \vect{\mathcal{B}}_\mathcal{X} \cdot \vect{\Sigma} \right] \right] }{8m_f^3} +\gX \gY  c_{\text{B}, \mathcal{XY}}^{(f)} \frac{ \vect{\mathcal{B}}_{\mathcal{X}} \cdot \vect{\mathcal{B}}_\mathcal{Y}}{8m_f^3} - \gX \gY c_{\text{E},\mathcal{XY}}^{(f)} \frac{\vect{\mathcal{E}}_\mathcal{X} \cdot \vect{\mathcal{E}}_\mathcal{Y}}{8m_f^3} \notag \\
    &  + \gX \gY \left[ \ii c_{\text{B}\Sigma, \mathcal{XY}}^{(f)} \frac{\vect{\mathcal{B}}_\mathcal{X} \times \vect{\mathcal{B}}_\mathcal{Y}  \cdot \vect{\Sigma}}{8m_f^3} - \ii c_{\text{E}\Sigma,\mathcal{XY}}^{(f)} \frac{\vect{\mathcal{E}}_\mathcal{X} \times \vect{\mathcal{E}}_\mathcal{Y} \cdot \vect{\Sigma}}{8m_f^3} + \cXIV \frac{ \text{Tr} \left\{ \mathcal{X}^{\mu \nu} \mathcal{Y}_{\mu \nu} \right\}}{4\Lambda^3} \right] \notag \\
    &  + \gX d_{D \mathcal{X}}^{(f)} \frac{\left[D^\mu, \mathcal{X}_{\mu i} \right] \gamma^i}{4m_f^2} - \gX \cIV \frac{\left[ D_t, \left\{ \ii \vect{D} \cdot, \vect{\mathcal{E}}_\mathcal{X} \right\} - \left\{ \vect{D} \times, \vect{\mathcal{E}}_\mathcal{X} \right\} \cdot \vect{\Sigma} \right]}{2 m_f \Lambda^2}, 
\end{align}
where we define the spin vector $\vect{\Sigma} = \begin{pmatrix}
    \vect{\sigma} & 0 \\
    0 & \vect{\sigma}
\end{pmatrix}$.
Apart from the last line, eq.~\eqref{Eq:L2} already has the familiar NR form known from NRQCD and NRQED. 
The corresponding NR Wilson coefficients~\cite{Hill2013} are summarized in table~\ref{tab:2} and are entirely determined by relativistic (or equivalently canonical) Wilson coefficients, where we also define $\bar{c}_{\mathcal{XY}\Sigma}^{(f)}= c_{\mathcal{XY}\sigma}^{(f)} + c_{D^2 \mathcal{X}}^{(f)} + c_{D^2 \mathcal{Y}}^{(f)}$.

\begin{table}[ht]
     \caption{Non-relativistic Wilson coefficients $c_\text{NR}^{(f)}$ in terms of relativistic Wilson coefficients $c_\text{rel}^{(f)}$.}
    \setlength{\tabcolsep}{0.0pt}
    \setlength{\extrarowheight}{2pt}
    \centering
    \begin{tabular}{ c  >{\hspace{1pt}} c <{\hspace{1pt}} ||  >{\hspace{1pt}} c    c }
    \toprule
      $c_\text{NR}^{(f)}$ & \multicolumn{1}{l||}{$= \hspace{1cm} c_\text{NR}^{(f)} \left( c_\text{Rel}^{(f)} \right)$}  & $c_\text{NR}^{(f)}$ & \multicolumn{1}{l}{$= \hspace{3cm} c_\text{NR}^{(f)} \left( c_\text{Rel}^{(f)} \right)$}  \\
    \midrule
    \rowcolor{gray!30}
       \rule{0pt}{1.8em} $c_{\text{F},\mathcal{X}}^{(f)}$ & $1+ \frac{m_f}{\Lambda} \cII $ & $c_{\text{W},\mathcal{X}}^{(f)}$ & $\frac{1}{2} \frac{m_f}{\Lambda} \cII +4  \frac{m_f^2}{\Lambda^2} \bar{c}_{D\mathcal{X}}^{(f)}  + 4\frac{m_f^3}{\Lambda^3} \cXI$  \\ [0.7em]
       \rule{0pt}{1.8em} $c_{\text{D},\mathcal{X}}^{(f)}$ & $1+ 2 \frac{m_f}{\Lambda} \cII + 8 \frac{m_f^2}{\Lambda^2} \bar{c}_{D\mathcal{X}}^{(f)}$ & $c_{\text{B},\mathcal{XY}}^{(f)}$ & $1 + 4 \frac{m_f^3}{\Lambda^3} \bar{c}_{\mathcal{XY}}^{(f)}$  \\ [0.7em]
       \rowcolor{gray!30}
       \rule{0pt}{1.8em} $c_{\text{S},\mathcal{X}}^{(f)}$ & $1+ 2 \frac{m_f}{\Lambda} \cII$ & $c_{\text{E},\mathcal{XY}}^{(f)}$ & $\bigg[ 1 + \frac{m_f \cII}{\Lambda} \bigg] \bigg[ 1+ \frac{m_f c_{\mathcal{Y}}^{(f)}}{\Lambda} \bigg] + 8 \frac{m_f^2}{\Lambda^2} \frac{\bar{c}_{D\mathcal{X}}^{(f)} + \bar{c}_{D\mathcal{Y}}^{(f)}}{2}  + 4 \frac{m_f^3}{\Lambda^3} \bar{c}_{\mathcal{XY}}^{(f)}$  \\ [0.7em]
       \rule{0pt}{2em} $c_{p^\prime p,\mathcal{X}}^{(f)}$ & $\frac{m_f}{\Lambda} \cII$ & $c_{\text{B}\Sigma,\mathcal{XY}}^{(f)}$ & $1+ \frac{m_f}{\Lambda} \frac{\cII+ c_\mathcal{Y}^{(f)}}{2} +8 \frac{m_f^2}{\Lambda^2} \frac{ \bar{c}_{D\mathcal{X}}^{(f)} + \bar{c}_{D\mathcal{Y}}^{(f)}}{2} +  4 \frac{m_f^3}{\Lambda^3}\bar{c}_{\mathcal{XY}\Sigma}^{(f)} $  \\ [0.7em]
       \rowcolor{gray!30}
       \rule{0pt}{2em} $c_{\text{M},\mathcal{X}}^{(f)}$ & $\frac{1}{2} \frac{m_f}{\Lambda} \cII+4 \frac{m_f^2}{\Lambda^2} \bar{c}_{D\mathcal{X}}^{(f)}$ & $c_{\text{E}\Sigma,\mathcal{XY}}^{(f)}$ & $\bigg[ 1 + \frac{m_f \cII}{\Lambda} \bigg] \bigg[ 1+ \frac{m_f c_{\mathcal{Y}}^{(f)}}{\Lambda} \bigg]+ 8 \frac{m_f^2}{\Lambda^2} \frac{ \bar{c}_{D\mathcal{X}}^{(f)} + \bar{c}_{D\mathcal{Y}}^{(f)}}{2} + 4 \frac{m_f^3}{\Lambda^3} \bar{c}_{\mathcal{XY}\Sigma}^{(f)} $  \\ [0.7em]
    \bottomrule
    \end{tabular}
    \label{tab:2}
\end{table}

At this order, the remaining two operations in figure~\ref{fig:FWroadmap} commute.
Therefore, we first eliminate the last term in eq.~\eqref{Eq:L2} through a fermion-field redefinition following our established scheme (red part in figure~\ref{fig:FWroadmap}) without affecting lower-order terms.
The other term with the new Wilson coefficient $d_{D\mathcal{X}}^{(f)}=1 + \frac{m_f}{\Lambda} \cII + 4 \frac{m_f^2}{\Lambda^2} \cIV$ in the last line of eq.~\eqref{Eq:L2} is converted into four-fermion terms through the gauge-field redefinition (purple part in figure~\ref{fig:FWroadmap}), in particular through
\begin{subequations}
    \begin{align}
        A_{\mathcal{F},i} =& A^{\prime}_{\mathcal{F},i} - e \sum_{f} Z_f  d_{D\mathcal{F}}^{(f)} \frac{ (\bar{f}^\prime \gamma_i f^\prime)}{4m_f^2}, \\
        \mathcal{A}_{\mathcal{G},i} =& \mathcal{A}^{\prime}_{\mathcal{G},i}  - g_s T^a \sum_{f} d_{D\mathcal{G}}^{(f)} \frac{ (\bar{f}^\prime T^a \gamma_i f^\prime)}{4m_f^2}.
    \end{align}
\end{subequations}

This field redefinition induces only two terms that are relevant in our order, which are
\begin{align}
    \Delta \mathcal{L} =& - \sum_{f} \bar{f}^\prime \gX d_{D\mathcal{X}}^{(f)} \frac{\left[ D^\mu, \mathcal{X}_{\mu i} \right] \gamma^i}{4m_f^2} f^\prime + \sum_{f, g}  \left[ g_\mathcal{X}^2 \frac{D_t \left( \bar{f}^\prime d_{D\mathcal{X}}^{(f)} T_\mathcal{X}^{a} \gamma^i  f^\prime \right) D_t \left( \bar{g}^\prime d_{D\mathcal{X}}^{(g)} T_\mathcal{X}^a \gamma^i  g^\prime \right)}{32 m_f^2 m_g^2}  \right] 
\end{align}
and follow from eq.~\eqref{Eq:Gauge}.
Even though this four-fermion term is suppressed by $m_f^{-2} m_g^{-2}$, the NR counting introduced in section~\ref{Sec:CR} implies that a time derivative acting on fermion fields effectively contributes an additional mass factor.\footnote{Like $f$, the index $g$ is a summation index labeling the fermion species and does not necessarily represent a gluon.}
Consequently, the term scales as $m_f^{-1} m_g^{-1}$ and will be addressed in section~\ref{Sec:4ferm}.

As a result, the two-fermion sector becomes diagonal and we reproduce all operators of NRQCD and NRQED to this order.
Within the algebraic NR reduction, the NR Wilson coefficients are determined by the short-distance information already encoded in the relativistic Wilson coefficients.
Provided that the relativistic matching input from the relativistic level includes the hard contributions associated with the subsequent scale separation~\cite{Labelle1998} between the relativistic and NR regime, all one-gauge-field terms in the NR Lagrangian ($c_{\text{F},\mathcal{X}}^{(f)}$, $c_{\text{D},\mathcal{X}}^{(f)}$, $c_{\text{S},\mathcal{X}}^{(f)}$, $c_{p^\prime p,\mathcal{X}}^{(f)}$,  $c_{\text{M},\mathcal{X}}^{(f)}$, and $c_{\text{W},\mathcal{X}}^{(f)}$) match their counterparts known from NRQCD and NRQED in terms of form factors mentioned in section~\ref{Sec:2}.
For two-gauge-field interactions, we find the NR Wilson coefficient $c^{(f)}_{\text{B},\mathcal{X}\mathcal{Y}} = 1+4 m_f^3 \bar{c}^{(f)}_{\mathcal{X} \mathcal{Y}}/\Lambda^3$, from which we can identify that the non-Born contribution is encoded in $\bar{c}_{\mathcal{XY}}^{(f)}$ after subtracting the Born contribution ($F_1(0)=1$).
In NRQCD and NRQED matching, the Born contribution denotes the part of the two-gauge-field amplitude fixed entirely by the elastic one-gauge-field form factors, whereas the non-Born part contains independent two-gauge-field structure information.
Isolating in $c_{\text{E},\mathcal{X}\mathcal{Y}}^{(f)}$ again $\bar{c}_{\mathcal{XY}}^{(f)}$ as the non-Born contribution, we obtain the Born contributions~\cite{Hill2013} encoded through form factors.\footnote{The relations of our NR Wilson coefficients compared to ref.~\cite{Manohar1997} are $c_{A1}^{(f)} = c_{\text{B},\mathcal{GG}}^{(f)}$, $c_{A1}^{(f)} +c_{A2}^{(f)}/2=c_{\text{E},\mathcal{GG}}^{(f)}$, $c_{A3}^{(f)} = 4 c_{ \left\{ \mathcal{GG} \right\} }^{(f)}$, and $c_{A4}^{(f)}= 0$ because the NR reduction does not generate new trace-like contributions.}
Similarly, we find the other two terms unique to NRQCD, namely $\vect{\mathcal{E}} \times \vect{\mathcal{E}} \cdot \vect{\Sigma}$ and $\vect{\mathcal{B}} \times \vect{\mathcal{B}} \cdot \vect{\Sigma}$, together with their NR Wilson coefficients $c_{\text{E}\Sigma,\mathcal{XY}}^{(f)}$ and $c_{\text{B}\Sigma,\mathcal{XY}}^{(f)}$.
These coefficients are available in the literature at tree level~\cite{Manohar1997}.
In addition, treating QCD and QED on equal footing allows us to find the NR Wilson coefficients corresponding to gluon-photon interactions.
This result shows that the derivation of the NR limit from a relativistic Lagrangian automatically encodes the hidden Lorentz~\cite{Hill2013} invariance, which emerges naturally from reorganizing the relativistic Lagrangian through field redefinitions.

\subsubsection{Non-relativistic four-fermion sector} \label{Sec:4ferm}

To diagonalize the four-fermion sector, we first extract mass factors from the time derivative term in $\Delta \mathcal{L}$ with the following field redefinition 
\begin{align}
    f^\prime = \left( 1- \ii g_\mathcal{X}^2 d_{D\mathcal{X}}^{(f)} T_\mathcal{X}^a \alpha^i  \sum_{g} \frac{D_t \left( \bar{g}^{\prime \prime} d_{D\mathcal{X}}^{(g)} T_\mathcal{X}^a \gamma^0 \alpha^i  g^{\prime \prime} \right)}{32m_f^2 m_g^2} + g_\mathcal{X}^2 d_{D\mathcal{X}}^{(f)} T_\mathcal{X}^a \alpha^i  \sum_{g} \frac{ \bar{g}^{\prime \prime} d_{D\mathcal{X}}^{(g)} T_\mathcal{X}^a \alpha^i  g^{\prime \prime}}{16m_f^2 m_g} \right) f^{\prime \prime}.
\end{align}
Here, we only display the linear contribution, since higher-order terms do not contribute at the order considered.
The field redefinition generates a $\Delta_\text{time}$ contribution, which cancels all remaining time derivative terms in the Lagrangian, and the corresponding $\Delta_\text{mass}$ part yields one relevant term at the order $m_f^{-1} m_g^{-1}$.
After this field redefinition, the Lagrangian takes the form $\mathcal{L} = \mathcal{L}_2 + \mathcal{L}_4 + \mathcal{L}_\text{Gauge} $, where $\mathcal{L}_2$ is the same as given by eq.~\eqref{Eq:L2} without the last two terms.
Abbreviating $\mathcal{I}^{\Gamma}_{f^{\prime \prime} g^{\prime \prime}, \mathcal{X}} = \left( \bar{f}^{\prime \prime} T^a_\mathcal{X} \Gamma f^{\prime \prime} \right) \left( \bar{g}^{\prime \prime} T^a_\mathcal{X} \Gamma g^{\prime \prime} \right)$, the four-fermion sector takes the form
\begin{align} \label{Eq:fft}
    \mathcal{L}_4 = \sum_{f, g}& \Bigg\{ d_{1,\mathcal{X}}^{(fg)} \frac{\mathcal{I}^{\mathds{1}}_{f^{\prime \prime} g^{\prime \prime}, \mathcal{X}}}{\Lambda^2}+ d_{\gamma,\mathcal{X}}^{(fg)} \frac{ \mathcal{I}^{\gamma^0}_{f^{\prime \prime} g^{\prime \prime}, \mathcal{X}} }{\Lambda^2} - d_{\gamma \gamma_5, \mathcal{X}}^{(fg)} \frac{ \mathcal{I}^{\gamma^0 \vect{\Sigma}}_{f^{\prime \prime} g^{\prime \prime}, \mathcal{X}}}{\Lambda^2} +2d_{\sigma,\mathcal{X}}^{(fg)} \frac{\mathcal{I}^{\vect{\Sigma}}_{f^{\prime \prime} g^{\prime \prime}, \mathcal{X}} }{\Lambda^2} + d_{\gamma_5,\mathcal{X}}^{(fg)} \frac{ \mathcal{I}^{\gamma_5}_{f^{\prime \prime} g^{\prime \prime}, \mathcal{X}} }{\Lambda^2}\notag \\
    & + d_{\gamma \gamma_5,\mathcal{X}}^{(fg)} \frac{ \mathcal{I}^{\gamma^0 \gamma_5}_{f^{\prime \prime} g^{\prime \prime}, \mathcal{X}} }{\Lambda^2}  -  \bar{d}_{\gamma,\mathcal{X}}^{(fg)} \frac{ \mathcal{I}^{\gamma^0 \vect{\alpha}}_{f^{\prime \prime} g^{\prime \prime}, \mathcal{X}}}{\Lambda^2}+ \left( - g_\mathcal{X}^2 d^{(f)}_{D\mathcal{X}} d^{(g)}_{D\mathcal{X}} +16 \frac{m_f m_g}{\Lambda^2} d_{\sigma,\mathcal{X}}^{(fg)} \right) \frac{ \mathcal{I}^{\vect{\alpha}}_{f^{\prime \prime} g^{\prime \prime}, \mathcal{X}} }{8 m_f m_g}   \Bigg\}
\end{align}
with the canonical Wilson coefficients $\bar{d}_{\gamma,\mathcal{X}}^{(fg)} = d_{\gamma,\mathcal{X}}^{(fg)} + \frac{m_f+m_g}{\Lambda} d_{5,\mathcal{X}}^{(fg)}$ and $\bar{d}_{\gamma,\mathcal{X}}^{(ff)}=0$.
Here, we do not display any odd four-fermion terms of the form $\left( \bar{f} \off f \right) \left( \bar{g} \dia g \right)$, \ie all terms proportional to $\Lambda^{-3}$ in eq.~\eqref{Eq:scattering3}, since they can be shifted to higher orders by redefining $f = \left( 1+ \frac{\off}{2m_f} \sum_g \left( \bar{g} \dia g \right)  + \gamma^0 \dia \sum_g \left( \bar{g} \frac{\gamma^0 \off}{2m_f}  g \right) \right) f^\prime$ without introducing new time derivatives acting on fermion fields in higher orders.
Only the last four terms in eq.~\eqref{Eq:fft} are not even and have to be diagonalized.
The structure of the two terms is $c_1 \mathcal{I}^{\Gamma}_{f g, \mathcal{X}} + c_2 \mathcal{I}^{\gamma^0 \Gamma}_{f g, \mathcal{X}}$ with $\Gamma= \gamma_5, \vect{\alpha}$.
Such a combination can be diagonalized by the field redefinition $f = \left( 1+ \frac{c_1 + c_2}{4m_f} \left[T_\mathcal{X}^a \gamma^0 \Gamma \sum_{g} \left( \bar{g}^\prime T_\mathcal{X}^a \gamma^0 \Gamma g^\prime \right) + T_\mathcal{X}^a \Gamma \sum_{g} \left( \bar{g}^\prime T_\mathcal{X}^a \Gamma g^\prime \right) \right] \right) f^\prime$, where again a linear redefinition suffices at the order considered.
The field redefinition is constructed such that the corresponding $\Delta_\text{time} =0$ part does not introduce new time derivatives on fermion fields and $\Delta_\text{mass}$ modifies the expression to $\frac{c_1-c_2}{2} \left( \mathcal{I}^{\Gamma}_{f^{ \prime} g^{ \prime}, \mathcal{X}} - \mathcal{I}^{\gamma^0 \Gamma}_{f^{ \prime} g^{\prime}, \mathcal{X}} \right)$.
The resulting four-fermion Lagrangian
\begin{align}
    \mathcal{L}_4 = \sum_{f,g} \Bigg\{& d_{1,\mathcal{X}}^{(fg)} \frac{\mathcal{I}^{\mathds{1}}_{f g, \mathcal{X}} }{\Lambda^2} + d_{\gamma,\mathcal{X}}^{(fg)} \frac{ \mathcal{I}^{\gamma^0}_{f g, \mathcal{X}} }{\Lambda^2} - d_{\gamma \gamma_5, \mathcal{X}}^{(fg)} \frac{ \mathcal{I}^{\gamma^0 \vect{\Sigma}}_{f g, \mathcal{X}} }{\Lambda^2}  + \frac{ d_{\gamma_5,\mathcal{X}}^{(fg)} - d_{\gamma \gamma_5,\mathcal{X}}^{(fg)}}{2} \frac{ \mathcal{I}^{\gamma^0}_{f g, \mathcal{X}} - \mathcal{I}^{\gamma^0 \gamma_5}_{f g, \mathcal{X}} }{\Lambda^2}    \notag \\
    & + 2  d_{\sigma,\mathcal{X}}^{(fg)} \frac{ \mathcal{I}^{\vect{\Sigma}}_{f g, \mathcal{X}} }{\Lambda^2}+ \frac{ g_\mathcal{X}^2 d^{(f)}_{D\mathcal{X}} d^{(g)}_{D\mathcal{X}} +16 \frac{m_f m_g}{\Lambda^2} d_{\sigma,\mathcal{X}}^{(fg)} + 8 \frac{m_f m_g}{\Lambda^2} \bar{d}_{\gamma,\mathcal{X}}^{(fg)} }{2} \frac{ \mathcal{I}^{\vect{\alpha}}_{f g, \mathcal{X}} - \mathcal{I}^{\gamma^0 \vect{\alpha}}_{f g, \mathcal{X}} }{8m_f m_g}\Bigg\},
\end{align}
with primes suppressed, does not yet appear diagonal.
However, the next section~\ref{Sec:NRLagrangian} shows that it can be cast into a diagonal form with respect to the upper and lower components of the Dirac four-spinor, but only when differences of the form $\mathcal{I}^{\Gamma}_{f^{ \prime} g^{ \prime}, \mathcal{X}} - \mathcal{I}^{\gamma^0 \Gamma}_{f^{ \prime} g^{\prime}, \mathcal{X}}$ enter the four-fermion sector.

\subsection{NR Lagrangian} \label{Sec:NRLagrangian}

We cast the Lagrangian into a completely diagonal form by separating $f = ( \phi_f, \chi_f)$ into upper $\phi_f$ and lower $\chi_f$ components, denoting two-component Pauli spinors.
With the help of the completeness relation~\cite{Haber2021} of $SU(N)$ generators ${(T^a)^{i}}_j {(T^a)^{k}}_{\ell} = \frac{1}{2} \left( {\delta^{i}}_{\ell} {\delta^{j}}_{k} - \frac{1}{N} {\delta^{i}}_{j} {\delta^{k}}_{\ell} \right)$ we can derive Fierz rearrangements~\cite{Pineda1998a} in the compact form
\begin{align}
\begin{split}
        &c_{1,\mathcal{X}}^{(fg)} \left( \phi_f^\dagger T_\mathcal{X}^{a} \chi_f \right) \left( \chi_g^\dagger T_\mathcal{X}^{a} \phi_g \right) + c_{2,\mathcal{X}}^{(fg)} \left( \phi_f^\dagger T_\mathcal{X}^{a} \sigma^i\chi_f \right) \left( \chi_g^\dagger T_\mathcal{X}^{a} \sigma^i \phi_g \right) \\
        =&- \tau_{\mathcal{X}\mathcal{Y}}^{(fg)} \left( c_{1,\mathcal{Y}}^{(fg)} + 3 c_{2,\mathcal{Y}}^{(fg)} \right) \left( \phi_f^\dagger T_\mathcal{X}^{a} \phi_g \right) \left( \chi_g^\dagger  T_\mathcal{X}^{a} \chi_f \right) \\
        &- \tau_{\mathcal{X}\mathcal{Y}}^{(fg)} \left( c_{1,\mathcal{Y}}^{(fg)} - c_{2,\mathcal{Y}}^{(fg)} \right) \left( \phi_f^\dagger T_\mathcal{X}^{a} \sigma^i \phi_g \right) \left( \chi_g^\dagger   T_\mathcal{X}^{a} \sigma^i\chi_f \right)  
\end{split}
\end{align}
for anti-commuting spinors and arbitrary coefficients $c_{1,\mathcal{X}}^{(fg)}$ and $c_{2,\mathcal{X}}^{(fg)}$.
The two representations are connected through $\tau_{\mathcal{X} \mathcal{Y}}^{(fg)}$ taking values $\tau_{\mathcal{F} \mathcal{F}}^{(fg)} = 1/\left( 2 N_{f,g} \right)$, $\tau_{\mathcal{G} \mathcal{F}}^{(fg)} = Z_f Z_g$, $\tau_{\mathcal{F} \mathcal{G}}^{(fg)} = \left( N_{f,g}^2 -1 \right)/\left( 4 Z_f Z_g N_{f,g}^2 \right)$, and $\tau_{\mathcal{G} \mathcal{G}}^{(fg)} = -1/\left( 2 N_{f,g} \right)$, where $N_{f,g}=3$ if $f$ and $g$ are both quark fields, and $N_{f,g}=1$ otherwise.\footnote{For the case of two quarks, color indices are contracted alongside spinor indices indicated by brackets around the spinors, but the case of a lepton-quark combination implies $\left( \phi^\dagger_f \phi_g \right)\left( \chi^\dagger_g \chi_f \right)=\left( \phi^\dagger_{f} \phi_{g,\alpha} \right)\left( \chi^\dagger_{g,\alpha} \chi_f \right)$.}
In addition, we introduce the charge-conjugated spinor $\chi_f^{(\text{C})}$.
For leptons that couple only through QED interactions, this spinor is defined via $\chi_f^{(\text{C})}= C_\mathcal{F} \chi_f^*$, where the charge-conjugation matrix is given by $C_\mathcal{F} \chi_f = \begin{pmatrix}
    0 & 1 \\
    -1 & 0
\end{pmatrix} \chi_f $.
For quarks, a corresponding matrix can be introduced to find the color-charge conjugated spinor, such that $\left( \chi^\dagger_f T^a_\mathcal{X} \chi_f \right) = \left( {\chi_f^{(\text{C})}}^\dagger \left( - T_\mathcal{X}^a \right)^{\text{T}} \chi_f^{(\text{C})} \right)$ holds, for example.
Applying these two manipulations, the NR Lagrangian reduces to
\begin{align} \label{Eq:NRQCED}
    \mathcal{L} = \sum_{f} \phi_f^\dagger \Bigg[& \ii D_t - m_f + \frac{\vect{D}^2}{2m_f} + \frac{\vect{D}^4}{8m_f^3} + \gX \left[ \frac{ \left\{ \vect{D}^2, \vect{\mathcal{B}}_\mathcal{X} \cdot \vect{\sigma} \right\} }{8m_f^3} +  c_{\text{F}, \mathcal{X}}^{(f)} \frac{\vect{\mathcal{B}}_\mathcal{X} \cdot \vect{\sigma}}{2m_f} +  c_{\text{D},\mathcal{X}}^{(f)} \frac{\left[ \vect{D} \cdot, \vect{\mathcal{E}}_\mathcal{X} \right]}{8m_f^2} \right]  \notag \\
    &+ \gX \left[ c_{\text{S},\mathcal{X}}^{(f)} \frac{ \left[ \ii \vect{D} \times, \vect{\mathcal{E}}_\mathcal{X} \right]\cdot \vect{\sigma}}{ 8m_f^2 } + c_{p^\prime p, \mathcal{X}}^{(f)} \frac{ \left\{ \vect{D} \cdot \vect{\sigma}, \vect{\mathcal{B}}_\mathcal{X} \cdot \vect{D} \right\} }{8m_f^3} + c_{\text{M},\mathcal{X}}^{(f)} \frac{ \left\{ \ii \vect{D} \cdot, \left\{ \vect{D} \times, \vect{\mathcal{B}}_\mathcal{X} \right\} \right\} }{8m_f^3} \right] \notag \\
    &  + \gX c_{\text{W},\mathcal{X}}^{(f)} \frac{ \left[ D_i , \left[ D_i, \vect{\mathcal{B}}_\mathcal{X} \cdot \vect{\sigma} \right] \right] }{8m_f^3} +\gX \gY  c_{\text{B}, \mathcal{XY}}^{(f)} \frac{ \vect{\mathcal{B}}_{\mathcal{X}} \cdot \vect{\mathcal{B}}_\mathcal{Y}}{8m_f^3} - \gX \gY c_{\text{E},\mathcal{XY}}^{(f)} \frac{\vect{\mathcal{E}}_\mathcal{X} \cdot \vect{\mathcal{E}}_\mathcal{Y}}{8m_f^3}  \notag \\
    & + \gX \gY \Bigg[ \ii \frac{c_{\text{B}\Sigma, \mathcal{XY}}^{(f)} \vect{\mathcal{B}}_\mathcal{X} \times \vect{\mathcal{B}}_\mathcal{Y} -  c_{\text{E}\Sigma,\mathcal{XY}}^{(f)} \vect{\mathcal{E}}_\mathcal{X} \times \vect{\mathcal{E}}_\mathcal{Y}}{8m_f^3} \cdot \vect{\sigma}   + \cXIV \frac{ \text{Tr} \left\{ \mathcal{X}^{\mu \nu} \mathcal{Y}_{\mu \nu} \right\}}{4\Lambda^3}\Bigg] \Bigg] \phi_f \notag \\
    &\hspace{-1.25cm}+ \sum_{f,g} \Bigg\{ d_{a,\mathcal{X}}^{(fg)} \frac{ \mathcal{I}_{\phi_f \phi_g,\mathcal{X}}^\mathds{1} }{m_f m_g} +d_{b,\mathcal{X}}^{(fg)} \frac{ \mathcal{I}_{\phi_f \phi_g,\mathcal{X}}^{\vect{\sigma}} }{m_f m_g} \Bigg\} + \left( \phi_{f/g} \to \chi_{f/g}^{(\text{C})} , T^a_\mathcal{X} \to - \left( T^a_\mathcal{X} \right)^\text{T} \right)  \notag \\
    &\hspace{-1.25cm}+\sum_{f,g} \Bigg\{ d_{c,\mathcal{X}}^{(fg)}  \frac{  \mathcal{I}_{\phi_f \chi^{\left(C\right)}_g,\mathcal{X}}^\mathds{1} }{m_f m_g} + d_{d,\mathcal{X}}^{(fg)}  \frac{\mathcal{I}_{\phi_f \chi^{\left(C\right)}_g,\mathcal{X}}^{\vect{\sigma}} }{m_f m_g} + 2\tau_{\mathcal{X} \mathcal{Y}}^{(fg)} d_{s,\mathcal{Y}}^{(fg)}  \frac{ \left( \phi^\dagger_f T^a_\mathcal{X} \phi_g \right) \left( \chi^{\text{(C)}\dagger}_f T^a_\mathcal{X} \chi^{(\text{C})}_g\right) }{m_f m_g} \notag \\
    &\hspace{-0.2cm}+ 2\tau_{\mathcal{X} \mathcal{Y}}^{(fg)} d_{v,\mathcal{Y}}^{(fg)}  \frac{ \left( \phi^\dagger_f T^a_\mathcal{X} \sigma^i \phi_g \right) \left( \chi^{\text{(C)}\dagger}_f T^a_\mathcal{X} \sigma^i \chi^{(\text{C})}_g \right) }{m_f m_g} \Bigg\} + \mathcal{L}_\text{Gauge}.
\end{align}
Due to the charge-conjugated spinor, we have a transparent interpretation of $\chi^{(\text{C})}_f$ as the annihilation operator of an antiparticle of type $f$. 
The NR Lagrangian associated with the antiparticle is the same as the Lagrangian of the particle, apart from $T_\mathcal{X}^a \to - \left( T_\mathcal{X}^a \right)^{\text{T}}$, which means for QED that the antiparticle has opposite charge $-Z_f$ compared to the particle, as expected.
The operators in the NR four-fermion sector are associated with the NR Wilson coefficients
\begin{subequations}
    \begin{align}
        d_{a,\mathcal{X}}^{(fg)} =& \frac{m_f m_g}{\Lambda^2} \left[ d_{1,\mathcal{X}}^{(fg)} + (1-\delta_{fg} )  d_{\gamma,\mathcal{X}}^{(fg)}  \right], \qquad  d_{b,\mathcal{X}}^{(fg)} = - \frac{m_f m_g}{\Lambda^2} \left[ d_{\gamma\gamma_5,\mathcal{X}}^{(fg)} - 2  d_{\sigma,\mathcal{X}}^{(fg)} \right], \\
        d_{c,\mathcal{X}}^{(fg)} =& 2 \frac{m_f m_g}{\Lambda^2} \left[ d_{1,\mathcal{X}}^{(fg)} - (1-\delta_{fg} )  d_{\gamma,\mathcal{X}}^{(fg)}  \right],  \quad \hspace{0.19cm} d_{d,\mathcal{X}}^{(fg)} = -2 \frac{m_f m_g}{\Lambda^2} \left[ d_{\gamma\gamma_5,\mathcal{X}}^{(fg)} + 2  d_{\sigma,\mathcal{X}}^{(fg)} \right], \\
        d_{s,\mathcal{X}}^{(fg)} =& -\frac{3}{2} \frac{g_\mathcal{X}^2}{4} d_{D,\mathcal{X}}^{(f)} d_{D,\mathcal{X}}^{(g)} -\frac{m_f m_g}{\Lambda^2} \left[  d_{\gamma_5,\mathcal{X}}^{(fg)}  + 3 (1-\delta_{fg} ) \bar{d}_{\gamma,\mathcal{X}}^{(fg)} -  d_{\gamma \gamma_5, \mathcal{X}}^{(fg)} + 6 d_{\sigma,\mathcal{X}}^{(fg)} \right],  \\
        d_{v,\mathcal{X}}^{(fg)} =&- \frac{1}{2} \frac{g_\mathcal{X}^2}{4} d_{D,\mathcal{X}}^{(f)} d_{D,\mathcal{X}}^{(g)} - \frac{m_f m_g}{\Lambda^2} \left[ -  d_{\gamma_5,\mathcal{X}}^{(fg)} + (1-\delta_{fg} ) \bar{d}_{\gamma,\mathcal{X}}^{(fg)} + d_{\gamma \gamma_5, \mathcal{X}}^{(fg)} + 2  d_{\sigma, \mathcal{X}}^{(fg)} \right],
    \end{align}
\end{subequations}
where $d_{D\mathcal{X}}^{(f)}=1 + \frac{m_f}{\Lambda} \cII + 4 \frac{m_f^2}{\Lambda^2} \cIV$ also contains a tree-level contribution.
Based on the combinations of fields, the interactions corresponding to Wilson coefficients $d_{a,\mathcal{X}}^{(fg)}$, $d_{b,\mathcal{X}}^{(fg)}$, $d_{c,\mathcal{X}}^{(fg)}$, $d_{d,\mathcal{X}}^{(fg)}$ describe NR scattering.
Since NR scattering is still a dynamical process in the NR regime, all these Wilson coefficients are only the relativistic Wilson coefficients and there are no tree-level contributions from the FW NR reduction.
On the other hand, $d_{s,\mathcal{X}}^{(fg)}$ and $d_{v,\mathcal{X}}^{(fg)}$ belong to EFT terms forming a point vertex of four fields, where particle and antiparticle of type $g$ are annihilated and particle and antiparticle of type $f$ are created.
These operators represent pair-creation and annihilation processes, which are not dynamical in the NR EFT and therefore appear as local contact interactions.
For renormalizable QED or QCD, their tree-level matching contributions are proportional to $g_\mathcal{X}^2 \propto \alpha_\mathcal{X}$.
The numerical values of the NR Wilson coefficients agree with the literature at tree level~\cite{Pineda1998b}.
Even without a relativistic EFT, performing the NR reduction on the QED or QCD Lagrangian would correctly yield the NR tree-level contribution of pair creation processes.
Comparing the results at loop level requires a matching of the initial relativistic four-fermion Wilson coefficients $d_{\Gamma,\mathcal{X}}^{(fg)}$.
The same applies to terms like $ \left( \phi^\dagger_f T^a_\mathcal{X} \phi_f \right) \left( \phi^{\dagger}_g T^a_\mathcal{X} \phi_g\right) /(m_f m_g)$, which are connected to $\phi_f^\dagger g_\mathcal{X} \frac{\left[ \vect{D} \cdot , \vect{\mathcal{E}}_\mathcal{X}\right]}{8m_f^2} \phi_f$ through the EOM.
In the standard NRQED/NRQCD basis used for comparison here, this particular EOM-equivalent contribution is conventionally represented by the latter two-fermion operator.
For a relativistic EFT matched to renormalizable QED or QCD, the corresponding local four-fermion coefficients may start only at loop level, depending on the channel and operator, but they need not vanish in general.

\section{Applications} \label{Sec:6}

In section~\ref{Sec:3}, we demonstrated the feasibility of our NR reduction as an alternative to a direct construction on the NR level.
To underline the versatility of this approach, we discuss here two sample applications: First, introducing corrections to NR center-of-mass and relative coordinates of composite systems relies on the NR form of the Poincar\'e generators~\cite{Peskin1995,Brambilla2003}.
However, in NR EFTs Lorentz covariance is no longer manifest, and the boost generator associated with Lorentz symmetry is not directly accessible in the NR formulation.
In a purely bottom-up NR EFT formulation without recourse to the relativistic parent theory, the boost generator can instead be parametrized using the symmetries of the NR theory, with additional coefficients, which can be fixed by explicitly imposing the commutation relations of the Poincar\'e algebra~\cite{Brambilla2003, Janson2025} or using the Lagrangian that has to be invariant under the corresponding non-linear boost transformation~\cite{Berwein2019}.
By contrast, starting from a Lorentz-covariant formulation allows us to derive the NR boost generator directly; we demonstrate this explicitly at order $\Lambda^{-2}$.
Second, the procedure for obtaining the NR limit of the Lagrangian is not limited to QCD and QED interactions.
Whenever a meaningful NR limit or description exists, we should be able to derive it via the NR reduction.
We illustrate this by adding a simple BSM field and deriving the corresponding NR Lagrangian together with its hidden Lorentz invariance.
In particular, we perform the reduction for an axion field~\cite{Weinberg1978,Wilczek1978,Kim2010}, a well-motivated candidate for solving the strong $CP$ problem, up to order $\Lambda^{-2}$.

\subsection{Poincar\'e generators}

For simplicity, we consider the Lagrangian from eq.~\eqref{Eq:LTinv}, which is still Lorentz covariant and invariant with respect to the Poincar\'e symmetry up to the order $\Lambda^{-2}$.
The Lagrangian
\begin{align} \label{Eq:Lquad}
    \mathcal{L} = \sum_f \bar{f} &\left[ \ii \slashed{D} - m_f - \gX \cII \frac{\mathcal{X_{\mu\nu} \sigma^{\mu\nu}}}{4\Lambda} + \gX \cIV \frac{ \left[ D^\mu , X_{\mu \nu} \right] \gamma^\nu}{\Lambda^2}   \right] f + \sum_{f,g,\Gamma}  \frac{ d_{\Gamma,\mathcal{X}}^{(fg)} \mathcal{I}_{fg,\mathcal{X}}^\Gamma}{\Lambda^2 } + \mathcal{L}_\text{Gauge}
\end{align}
is then our starting point.
To derive the NR Poincar\'e generators, particularly the NR boost generator, we parametrize this Lagrangian in terms of fundamental fields $\mathcal{L} = \mathcal{L} \left( \bar{f}, f,\partial_\mu f, A_{\mathcal{X},\sigma} , \partial_\mu A_{\mathcal{X},\sigma} , \partial_\mu \partial_\nu A_{\mathcal{X},\sigma} \right)$, which transform under a general symmetry according to $\phi \left( x \right) \to \phi^\prime ( x) = \phi \left( x \right) + \delta \phi \left( x \right)$ for all involved fields.
The variation of the Lagrangian $\delta \mathcal{L}$ then yields the conserved Noether charge density
\begin{align} \label{Eq:charge}
    j^0 =& \sum_{f} \frac{\partial \mathcal{L}}{\partial \left( \partial_0 f \right)} \delta f + \left[ \frac{\partial \mathcal{L}}{\partial \left( \partial_0 A_{\mathcal{X},\sigma}^a \right)} - \partial_\nu \frac{\partial \mathcal{L}}{\partial \left( \partial_\nu \partial_0 A_{\mathcal{X},\sigma}^a \right)} \right] \delta A_{\mathcal{X},\sigma}^a + \frac{\partial \mathcal{L}}{\partial \left( \partial_0 \partial_\nu A_{\mathcal{X},\sigma}^a \right)} \partial_\nu \delta A_{\mathcal{X},\sigma}^a \notag \\
    &+ \delta x^0 \mathcal{L},
\end{align}
generalized to higher-order derivatives.
The presence of higher-order derivatives in the Lagrangian introduces a freedom in the choice of the order of partial integrations.
When we use a symmetrized partial integration, the order of partial derivatives can be treated interchangeably, \ie $\partial\left( \partial_\mu \partial_\nu \phi \right) / \partial \left( \partial_\alpha \partial_\beta \phi \right) = \partial\left( \partial_\mu \partial_\nu \phi \right) / \partial \left( \partial_\beta \partial_\alpha \phi \right)$.
In contrast, one may also perform partial integrations asymmetrically, in which case ordered derivatives are kept distinct.

The latter asymmetric choice yields manifestly gauge-invariant results, and we adopt this convention in the following.
The main task in deriving the NR boost generator is to properly handle higher-order time derivatives acting on gauge fields.
Because the boost generator will ultimately take the form $f^\dagger \left\{ \vect{x} , h \right\} f$, even unitary transformations used to remove even EFT terms with second-order time derivatives on gauge fields are shifted to lower orders due to the mass term $\gamma^0 m_f$ contained in the Hamiltonian $h$.
To ensure no higher-order terms feed into lower orders, we first remove the second-order time derivative acting on the gauge field via a unitary field redefinition $f= u f^\prime$ with $ u= u(\mathcal{X}_{0i}) = \exp{\ii \gX \cIV \frac{\vect{\mathcal{E}}_\mathcal{X} \cdot \vect{\alpha}}{\Lambda^2} }$.
This transformation can be understood as a canonical reduction of the gauge-field conjugate momentum.
Applying this redefinition in eq.~\eqref{Eq:charge} amounts to an application of the chain rule together with the on-shell conditions for the transformed fields $\bar{f}^\prime$, $f^\prime$, and yields
\begin{align} \label{Eq:cc}
    j^0 \cong \frac{\partial \mathcal{L}^\prime}{\partial \left( \partial_0 f^\prime \right)} u^\dagger \delta f + \left[ \frac{\partial \mathcal{L}^\prime}{\partial \mathcal{X}_{0j}} - \partial_i \frac{\partial \mathcal{L}^\prime}{\partial \left( \partial_i \mathcal{X}_{0j} \right)} \right] \delta \mathcal{A}_{\mathcal{X},j} + \delta x^0 \mathcal{L}^\prime + \frac{\partial \mathcal{L}^\prime}{\partial \left( \partial_0 f^\prime \right)} \frac{\partial f^\prime}{\partial \mathcal{X}_{0i}} \delta \mathcal{X}_{0i}.
\end{align}
Note that the equation holds only for the Lagrangian from eq.~\eqref{Eq:Lquad} and for $u$ only expanded to linear order, which suffices in our case. The benefit of eq.~\eqref{Eq:cc} is that we can use the redefined Lagrangian and take derivatives with respect to the new fields, while still using the simple transformation properties of the covariant field $f$ rather than with respect to the more involved expression $f^\prime$.
As a result, the conserved charges associated with the Poincar\'e symmetries consist of spacetime translations characterized by the transformation $\delta \phi = - a^\nu \partial_\nu \phi$ for all fields $\phi = f, A^a_{\mathcal{X},\sigma}$ together with $\delta x^\nu = a^\nu $, yielding $j^0 = - a_\nu T^{0\nu}$ such that the energy-momentum tensor $T^{0\nu}$ is conserved.
The Poincar\'e symmetries are completed by Lorentz transformations under which the fields transform as $\delta f = - \frac{\ii}{2} \omega_{\mu \nu} \left[ \ii \left( x^\mu \partial^\nu - x^\nu \partial^\mu \right) + \frac{\sigma^{\mu \nu}}{2} \right]f$ and $\delta A^a_{\mathcal{X},\sigma}= - \frac{\ii}{2} \omega_{\mu \nu} \left[ \ii \left( x^\mu \partial^\nu - x^\nu \partial^\mu \right) \delta^\rho_\sigma + {\left(\Sigma^{\mu \nu}\right)_\sigma}^\rho \right] A^a_{\mathcal{X},\rho}$ where ${\left(\Sigma^{\mu \nu}\right)_\sigma}^\rho= \ii \left({\delta^\mu}_\sigma \eta^{\nu \rho} - {\delta^\nu}_\sigma \eta^{\mu \rho} \right)$ are the gauge-field spin matrices.
These transformations, together with $\delta x^\nu = {\omega^\nu}_\mu x^\mu$, where $\omega_{\mu\nu} = - \omega_{\nu \mu}$, yield the corresponding conserved charge $j^0 = \frac{1}{2} \omega_{\mu\nu} J^{\mu \nu}$ corresponding to the generator of rotations and Lorentz boosts $J^{\mu \nu}$.
As long as gauge fields are on-shell (the EOM for $A^a_{\mathcal{X},0}$ holds), the manifestly gauge invariant generator densities
\begin{subequations} \label{Eq:Poigen}
    \begin{align} 
        \mathcal{T}^{0\nu} =& \pi_{f^\prime} D^\nu f^\prime + \pi_\mathcal{X}^j {\mathcal{X}^{\nu}}_{j} - \eta^{0\nu} \mathcal{L}^\prime \\
        \mathcal{J}^{\mu \nu} =& x^\mu \mathcal{T}^{0\nu} - x^\nu \mathcal{T}^{0\mu} - \pi_{f^\prime} \left[  \frac{\ii}{2} u^\dagger \sigma^{\mu \nu} u - \left( {\delta^{\mu}}_{i} {\mathcal{X}_0}^{\nu} - {\delta^{\nu}}_{i} {\mathcal{X}_0}^{\mu} + {\delta^{\mu}}_{0} {\mathcal{X}^\nu}_{i} - {\delta^\nu}_{0} {\mathcal{X}^{\mu}}_{i} \right) \frac{\partial u^\dagger}{\partial \mathcal{X}_{0i}} u \right] f^\prime
    \end{align}
\end{subequations}
follow.
Next, we separate eq.~\eqref{Eq:Poigen} into the 4 generator densities: Hamiltonian, momentum, angular momentum, as well as the boost generator, and insert the definition for $u= \exp{\ii \gX \cIV \frac{\vect{\mathcal{E}}_\mathcal{X} \cdot \vect{\alpha}}{\Lambda^2} }$ expanded until the linear order to find
\begin{subequations} \label{Eq:generators}
   \begin{align}
        \mathcal{T}^{00} = \mathcal{H} =& \sum_{f} f^{\prime \dagger} h f^\prime + \mathcal{H}_\text{Gauge} + \mathcal{H}_4 \\
        \mathcal{T}^{0i} = \mathcal{P}^i =& \sum_{f} f^{\prime \dagger} \ii D^i f^\prime + \frac{\left( \vect{\pi}_\mathcal{X} \times \vect{B}_\mathcal{X} \right)^i - \left( \vect{B}_\mathcal{X} \times \vect{\pi}_\mathcal{X} \right)^i }{2}  \\
       \frac{1}{2} \varepsilon^{ijk} \mathcal{J}^{jk} = \mathcal{J}^i =& \left( \vect{x} \times \vect{\mathcal{P}} \right)^i + \sum_f f^{\prime \dagger} \frac{\Sigma^i}{2}  f^\prime \\
        \mathcal{J}^{0i} = \mathcal{K}^i =& t \mathcal{P}^i - \sum_{f} f^{\prime \dagger} \left[ \frac{ \left\{ x^i , h \right\} }{2} + \gX \cIV \frac{ \left( \vect{\mathcal{B}}_\mathcal{X} \times \vect{\alpha} \right)^i}{\Lambda^2} \right] f^\prime - x^i \mathcal{H}_{\text{Gauge}} - x^i \mathcal{H}_4,
   \end{align}
\end{subequations}
where $h$, $\mathcal{H}_\text{Gauge}$, and $\mathcal{H}_4$ follow trivially from the Legendre transformation of the Lagrangian represented with respect to $f^\prime$. 

When we carry out the subsequent field redefinitions from the main part, \ie eq.~\eqref{Eq:Reduction}, we replace $\left(1+ \frac{m_f}{\Lambda} \cII \right) \to \left(1+ \frac{m_f}{\Lambda} \cII + 4 \frac{m_f^2}{\Lambda^2} \cIV \right)$ in the term $\vect{\mathcal{E}}_\mathcal{X}\cdot \vect{\alpha}$ due to the previous redefinition of the time derivative.
For these field transformations, one could apply the chain rule as before.
However, the additional advantage of working with the generators from eq.~\eqref{Eq:generators} is that they correspond to a canonical Lagrangian without higher-order conjugate momenta.
In fact, we can perform fermion field redefinitions directly at the level of the generators.
As the time derivative of the NRLO Lagrangian is exchanged with conjugate momenta when moving from the Lagrangian to the generators, we have to express the original gauge-field conjugate momentum $ \pi_\mathcal{X}^{a,i} = \frac{\partial \mathcal{L}}{\partial \left( \partial_t \vect{A}^a_{\mathcal{X},i} \right)}$ in terms of the new conjugate momentum instead, which is obtained from the redefined Lagrangian $\mathcal{L}^\prime$.
Alternatively, we could quantize the generators in eq.~\eqref{Eq:generators} and perform the redefinitions in the Hamiltonian picture, equivalent to the procedure described above.

Note that, after the field redefinitions, the NR Lagrangian is still covariant with respect to spacetime translations and rotations.
Consequently, the Hamiltonian, momentum and angular momentum coincide with those obtained by a naive derivation from the NR Lagrangian.
The boost generator, however, is modified in a nontrivial manner, as expected since covariance with respect to Lorentz boost is lost in the NR limit.
These nontrivial contributions result solely from calculating $u^\dagger \vect{x} u$ and from the changes in the conjugate momenta and vector potentials due to field redefinitions.
The NR boost generator takes the form
\begin{align}
    \vect{\mathcal{K}} = t \vect{\mathcal{P}} - \sum_{f} f^{\dagger} \left( \frac{ \left\{ \vect{x}, h_\text{NR} \right\} }{2} + \gamma^0 \frac{\vect{\Sigma} \times \ii \vect{D}}{2m_f} - \gX c_{\text{D},\mathcal{X}}^{(f)} \frac{\vect{\pi}_\mathcal{X}}{8m_f^2}\right) f - \vect{x} \mathcal{H}_{\text{Gauge}} - \vect{x} \mathcal{H}_{4,\text{NR}}.
\end{align}
Again, the trivial parts $h_\text{NR}$ and $\mathcal{H}_{4,\text{NR}}$ follow straightforwardly from the Legendre transformation of the NR Lagrangian and are therefore left implicit.
We recover the established~\cite{Heinonen2012,Brambilla2003} nontrivial extension at the order $m_f^{-1}$ for the boost generator density, which is a pure tree-level effect.
Beyond the universal $1/m_f$ contribution, the first radiative corrections enter at order $m_f^{-2}$ through $c_{\text{D},\mathcal{X}}^{(f)}$ that belongs to an electric-field term contained in the conjugate momentum $\vect{\pi}_\mathcal{X}$, in agreement with eq.~(54) in ref.~\cite{Berwein2019}.

\subsection{BSM extensions}
As an illustrative extension of the initial Lagrangian from eq.~\eqref{Eq:LTinv}, which includes only QCD and QED interactions, we add the EFT Lagrangian for an axion~\cite{Weinberg1978,Wilczek1978,Kim2010} and discuss the NR limit of the combined theory and consider all operators up to the order $\Lambda^{-2}$.
The axion Lagrangian is constructed from a pseudoscalar field $a$, \ie it is odd under parity ($P$) and time-reversal ($T$) symmetry.
The axion field is understood as a pseudo-Goldstone boson~\cite{Weinberg1972} of a spontaneously broken Peccei-Quinn $U(1)$ symmetry~\cite{Peccei1977,Peccei1977a}.
Hence, it exhibits an approximate shift symmetry.
Derivative fermion interactions respect the perturbative continuous shift symmetry, whereas the anomalous $aG\tilde G$ coupling together with nonperturbative QCD effects generates the axion potential and mass.
The leading-order Lagrangian is given by $\mathcal{L}_0 = \frac{1}{2} \left( \partial_\mu a \right) \left( \partial^\mu a \right) - \frac{1}{2} m_a^2 a^2+ \left( \theta_\mathcal{X} - c_{a,\mathcal{X}\tilde{\mathcal{X}}} \frac{a}{\Lambda} \right) \frac{g_\mathcal{X}^2}{32\pi^2} \mathcal{X}_{\mu \nu} \tilde{\mathcal{X}}^{\mu \nu}$, where $\theta_G$ denotes the strong-$CP$ angle. 
An analogous electromagnetic topological term may formally be included, although in ordinary QED it does not have the same local physical significance.

We restrict our analysis to EFT terms that are even under both $P$ and $T$.
Once effects from breaking the shift symmetry are retained, a general treatment also requires the additional non-derivative axion operators allowed by the remaining symmetries and power counting.
Motivated by the approximate shift symmetry of a QCD-like axion, we restrict the following discussion to higher-dimensional EFT terms that preserve $P$, $T$, and shift symmetry.
With these simplifications, the shift-symmetry-preserving axion EFT Lagrangian up to order $\Lambda^{-2}$ involves only bilinear fermion terms, represented by $\mathcal{L}_{a,2} =  \sum_{f} \bar{f} \left\{  c_{\partial a}^{(f)} \frac{\left( \partial_\mu a \right)}{\Lambda} \gamma^\mu \gamma_5  + c_{\partial^2 a}^{(f)} \frac{ \left( \partial^2 a \right)}{\Lambda^2} \ii \gamma_5  \right\} f$, where $\partial^2 = \partial_\mu \partial^\mu$.
Through the field redefinition $f = \left( 1- 2 c_{\partial a D}^{(f)} \frac{ \left( \partial_\mu a \right) }{\Lambda^2} \gamma^\mu \gamma_5 \right) f^\prime$, the other allowed term $c_{\partial a D}^{(f)} \varepsilon_{\mu \nu \alpha \beta} \frac{ \left\{ \ii D^\mu, \left( \partial^\nu a \right) \right\} }{\Lambda^2} \sigma^{\alpha \beta}$ in $\Lambda^{-2}$ has already been eliminated.
The relativistic axion EFT is specified through $\mathcal{O}(\Lambda^{-2})$ and, in the following, closed-form expressions generated by iterative field redefinitions retain higher powers of the Wilson coefficients as a partial resummation. 
Such terms should not be interpreted as a complete EFT prediction at $\mathcal{O}(\Lambda^{-3})$ or higher.
To prevent operators that have been shifted to higher orders by field redefinitions from re-entering lower orders through subsequent applications of the EOM of the massive axion, we rewrite the Lagrangian with respect to the LO axion EOM $\mathcal{K}_a = \left( \partial^2 a \right) + m_a^2 a$, as explained also in figure~\ref{fig:FWroadmap}.
Consequently, we perform the NR derivation on the Lagrangian
\begin{align}
    \mathcal{L}_{a,2} = \sum_f \bar{f} \left( - \frac{m_a^2}{\Lambda^2} c_{\partial^2 a}^{(f)} a \ii \gamma_5 + c_{\partial a}^{(f)} \gamma^0 \frac{a_0 \gamma_5 + \vect{a} \cdot \vect{\Sigma} }{\Lambda} + c_{\partial^2 a}^{(f)} \frac{ \mathcal{K}_a }{\Lambda^2} \ii \gamma_5  \right) f,
\end{align}
represented in terms of odd and even operators proportional to $a_0 \gamma_5= \left( \partial_0 a \right) \gamma_5$ and $ \vect{a} \cdot \vect{\Sigma} = \left( \partial_i a \right) \Sigma^i$, respectively.
Within this order, no axion-dependent four-fermion terms arise, such that $\mathcal{L}_{a,4}=0$.
 
Next, we carry out the NR reduction of the full Lagrangian $\mathcal{L}_{\text{QCED}}+\mathcal{L}_0 + \mathcal{L}_{a,2}$, where $\mathcal{L}_{\text{QCED}}$ refers to eq.~\eqref{Eq:LTinv}, neglecting orders $\Lambda^{-3}$.
As we already derived the diagonalization procedure of $\mathcal{L}_\text{QCED}$, we first apply these field redefinitions to find $\mathcal{L}_\text{NRQCED}$, which corresponds to the $\Lambda^{-2}$ contribution from eq.~\eqref{Eq:NRQCED}.
The same field redefinitions must also be applied to $\mathcal{L}_{a,2}$.
Note that due to the restriction to the order $\Lambda^{-2}$, there is no canonical reduction required, and we can directly apply the field redefinition $f = \exp{ \gamma_0  \ii \vect{D} \cdot \vect{\alpha} / (2 m_f )} \exp{-\ii \gY c_{\text{F},\mathcal{Y}}^{(f)}  \vect{\mathcal{E}}_\mathcal{Y} \cdot \vect{\alpha} / (4m_f^2)} f^\prime$ to obtain the modified Lagrangian
\begin{align} \label{Eq:axion2}
\begin{split}
    \mathcal{L}_{a,2} = \sum_f \bar{f}^{\prime} \Bigg(& - c_{\partial^2 a}^{(f)} \frac{m_a^2}{\Lambda^2} a \ii \gamma_5 + c_{\partial a}^{(f)} \gamma^0 \frac{a_0 \gamma_5}{\Lambda} + c_{\text{F},a}^{(f)} \gamma^0 \frac{ \vect{a} \cdot \vect{\Sigma} }{2m_f} -  c_{\text{S}, a}^{(f)} \frac{ \left\{ \ii \vect{D} \cdot \vect{\Sigma} , a_0 \right\} }{8 m_f^2} \\
    &- \gX c_{\text{EDM},\mathcal{X}}^{(f)} a \frac{\vect{\mathcal{E}}_\mathcal{X} \cdot \vect{\Sigma}}{2m_f^2} + c_{\partial^2 a}^{(f)} \frac{ \mathcal{K}_a }{\Lambda^2} \ii \gamma_5  \Bigg) f^\prime.
\end{split}
\end{align}
We defined the axion NR Wilson coefficients $c_{\text{F},a}^{(f)} = 2  \frac{m_f}{\Lambda} c_{\partial a}^{(f)} + \frac{m_a^2}{\Lambda^2} c_{\partial^2 a}^{(f)} $ parametrizing the \textit{axion-wind term}~\cite{Smith2024} where $\vect{a}$ couples to the spin, while $c_{\text{S},a}^{(f)} = 4 m_f c_{\partial a}^{(f)}/\Lambda$ corresponds to the \textit{axioelectric term} involving the coupling of $a_0$, as well as the NR Wilson coefficient $c_{\text{EDM},\mathcal{X}}^{(f)}= m_a^2 c_{\partial^2 a}^{(f)} c_{F,\mathcal{X}}^{(f)} /(2\Lambda^2)$ of the axion-induced NR electric-dipole moment (EDM).
For simplicity, we again omit the explicit field redefinitions that remove odd operators at order $m_f^{-2}$.

Before redefining the gauge fields and the axion field to shift odd time-derivative terms into the four fermion sector, we must address the first two odd terms in eq.~\eqref{Eq:axion2}.
Their diagonalization is not straightforward because the operators $\varepsilon_f a \ii \gamma_5/2$ with $\varepsilon_f =-2 c_{\partial^2 a}^{(f)} m_a^2/\Lambda^2$ and $\eta_f \gamma^0 a_0 \gamma_5/m_f$ with $\eta_f = m_f c_{\partial a}^{(f)}/\Lambda$ form a cycle under the FW field redefinitions:
Eliminating the $a \ii \gamma_5$ term (via a suitable $\Delta_\text{mass}$ contribution) generates a $a_0 \gamma^0 \gamma_5$ term through the corresponding $\Delta_\text{time}$ contribution of the field redefinition.
Eliminating the latter induces $\partial_0 a_0 \ii \gamma_5$, which, once again, implies the initial $a \ii \gamma_5$ term, once we rewrite $\partial_0 a_0 \ii \gamma_5$ through the EOM as $\mathcal{K}_a \ii \gamma_5$.
Consequently, the two operators are mapped into each other under this procedure and cannot be removed independently.
Hence, we exploit this fact and perform the field redefinition
\begin{align} \label{Eq.:axfr}
    f^\prime = \exp{ \frac{1}{2} \varepsilon_f \frac{a \ii \gamma_5}{2m_f} + \eta_f \frac{a_0 \gamma^0 \gamma_5}{2m_f^2}} \exp{ - \frac{1}{2} \eta_f \frac{m_a^2}{m_f^2} \frac{a \ii \gamma_5}{2m_f} -\frac{1}{4} \varepsilon_f \frac{a_0 \gamma^0 \gamma_5}{2m_f^2}} f^{\prime \prime}.
\end{align}
This choice treats both odd operators on equal footing, such that no particular ordering of their elimination is preferred.
The field redefinition, eq.~\eqref{Eq.:axfr},  yields a replacement of the first two terms in the Lagrangian, eq.~\eqref{Eq:axion2}, abbreviated by $\xi_f = \frac{1}{2} \varepsilon_f a \ii \gamma_5 + \eta_f \gamma^0 \frac{a_0 \gamma_5}{m_f}$ to
\begin{align}
     \xi_f  \to - \frac{1}{8} \left( \varepsilon^2_f + \left( \frac{m_a}{m_f} \right)^4 \eta_f^2 \right) \frac{a^2}{m_f} + \left(\frac{m_a}{2 m_f} \right)^2 \xi_f + \left( \eta_f - \frac{\varepsilon_f}{4} \right) \frac{\mathcal{K}_a \ii \gamma_5}{2m_f^2}.
\end{align}
Hence, after a single iteration, we generate new contributions, a mass term proportional to $a^2/m_f$ and a modification to the odd axion EOM proportional to $\mathcal{K}_a \ii \gamma_5$, while our original two operators abbreviated by $\xi_f$ are reproduced, but suppressed by a factor of $+ m_a^2/(2m_f)^2 $.
Repeating the procedure yields a sequence in which these two terms are successively suppressed, while the mass term and the odd axion EOM term accumulate into separate geometric series, which converge provided $m_a < 2m_f$.
For a QCD axion, $m_a \ll m_e$, and thus the Lagrangian for the axion is recast into $\mathcal{L}_{a,2}= \sum_f \bar{f}^{\prime \prime} \mathcal{I}_{f,a} f^{\prime \prime}$ with
\begin{align} \label{Eq:NRaxion2}
    \mathcal{I}_{f,a} = - c_m^{(f)}  \frac{a^2}{8m_f}   +c_{\text{F},a}^{(f)} \frac{\vect{a} \cdot \vect{\Sigma} }{2m_f}  - \frac{m_f}{\Lambda}c_{\partial a}^{(f)} \gamma^0 \frac{ \left\{ \ii \vect{D} \cdot \vect{\Sigma} , a_{0} \right\} }{4m_f^2} - \gX c_{\text{EDM},\mathcal{X}}^{(f)} a \frac{\vect{\mathcal{E}}_\mathcal{X} \cdot \vect{\Sigma}}{2m_f^2} + d_{\partial^2 a}^{(f)} \frac{ \mathcal{K}_a }{2m_f^2} \ii \gamma_5
\end{align}
the modified kernel of the Lagrangian.
This Lagrangian is parametrized through additional NR Wilson coefficients
\begin{subequations}
\begin{align}
    d_{\partial^2 a}^{(f)} &= \left[ m_f c_{\partial a}^{(f)}/\Lambda + 2 m_f^2 c_{\partial^2 a}^{(f)} /\Lambda^2 \right] \left[ 1- m_a^2 /(2m_f)^2 \right]^{-1} \\
    c_m^{(f)} &= \left[ \varepsilon_f^2 + \eta_f^2 m_a^4 / m_f^4 \right] \left[ 1 - m_a^4/(2m_f)^4 \right]^{-1}
\end{align}
\end{subequations}
 belonging to an odd axion LO EOM term and an axion-induced mass shift, respectively.
With these manipulations, the NR two-fermion Lagrangian including the axion sector is complete.

The last step is the redefinition of gauge fields together with a redefinition of the axion field $a = a^\prime + \sum_{f} \bar{f}^{\prime} d_{\partial^2 a}^{(f)} \frac{ \ii \gamma_5}{2m_f^2} f^\prime$, shifting the last term in eq.~\eqref{Eq:NRaxion2} to the four-fermion sector.
Applying the same diagonalization procedure as in the previous four-fermion sector, we obtain the additional axion-induced NR four-fermion Lagrangian
\begin{align}
\begin{split}
    \mathcal{L}_{a,4} =& - \sum_{f, g} d_{\partial^2 a}^{(f)} d_{\partial^2 a}^{(g)} \left( 1- \frac{m_a^2}{4m_f m_g} \right) \frac{ \left( \bar{f}^\prime \gamma^0 \gamma_5 f^\prime \right) \left( \bar{g}^\prime \gamma^0 \gamma_5 g^\prime \right) - \left( \bar{f}^\prime \gamma_5 f^\prime \right) \left( \bar{g}^\prime \gamma_5 g^\prime \right) }{4m_f m_g}  \\
    &= - \sum_{f,g} \frac{d_{\partial^2 a}^{(f)} d_{\partial^2 a}^{(g)}}{2m_f m_g} \left( 1- \frac{m_a^2}{4m_f m_g} \right) \frac{2 \tau^{(fg)}_{\mathcal{FX}}}{Z_f Z_g}  \\
    &\hspace{0.5cm}\times\Bigg\{ \left( \phi^\dagger_f T^a_\mathcal{X} \phi_g \right) \left( {\chi_f^{\text{(C)}}}^\dagger T^a_\mathcal{X} \chi^{(\text{C})}_g\right)-  \left( \phi^\dagger_f T^a_{\mathcal{X}} \sigma^i \phi_g \right) \left( {\chi_f^{\text{(C)}}}^\dagger T^a_{\mathcal{X}} \sigma^i \chi^{(\text{C})}_g \right) \Bigg\}.
\end{split}
\end{align}

As a result, the NR Lagrangian is given by $\mathcal{L} = \mathcal{L}_\text{NRQCED} + \mathcal{L}_{a,0} + \mathcal{L}_{a,2} + \mathcal{L}_{a,4}$.
Our derivation yields a set of axion-induced NR operators together with the NR Wilson coefficients, in which the relations implied by the underlying Lorentz invariance are automatically encoded.
In particular, we find that the axion mass generates an $a^2$-dependent scalar interaction, which can be interpreted as an axion-background-dependent correction to the fermion mass, and induces an NR EDM-like interaction.
Furthermore, the tree-level axion-spin coupling ($\vect{a} \cdot \vect{\Sigma}$) receives axion-mass dependent corrections, and we obtain axion-induced NR pair-creation terms. 
In accordance with the literature~\cite{Berlin2024}, the EDM term vanishes if we only include the leading-order axion-fermion coupling in the limit $c_{\partial^2 a}^{(f)} \to 0$.
The influence of the axion's mass term on the NR reduction is usually not addressed.

\section{Discussion} \label{Sec:5}

Our work lies at the interface between relativistic and NR EFTs, as well as their connection through the FW transformation.
In this section, we do not provide a comprehensive review of the extensive literature on FW transformations and NR reductions, but instead provide a selective overview to position our approach and facilitate comparison with existing methods.

The FW transformation~\cite{Foldy1950,Tani1951} was originally developed for the Dirac Hamiltonian of a spin-$1/2$ particle interacting with an external electromagnetic field, with the aim of block-diagonalizing the Hamiltonian with respect to upper and lower components of the Dirac four-spinor via an iterative sequence of unitary transformations performed order-by-order.
In this way, both the electron and positron degrees of freedom decouple in the Dirac equation, and the description reduces to the NR Schrödinger-Pauli description, supplemented by relativistic corrections.
This approach is widely used and remains an active tool for deriving higher-order NR expansions of relativistic single-particle Hamiltonians~\cite{Jentschura2024}.

Besides its original formulation, the FW transformation has found applications in heavy-quark systems in quantum chromodynamics~\cite{Balk1994}, relativistic quantum chemistry~\cite{Douglas1974,Hess1986,Wolf2002}, extensions of the Standard Model~\cite{Smith2024}, and fermions propagating in non-inertial spacetime backgrounds~\cite{Obukhov2009,Obukhov2013}, where it has been used to derive NR limits and to identify post-Newtonian gravitational interactions~\cite{Silenko2005}. 
Furthermore, the formalism has been generalized from spin-$1/2$ to arbitrary-spin particles~\cite{Silenko2025} and has also been applied in field-theoretical treatments in both Hamiltonian~\cite{Dong1977} and Lagrangian~\cite{Koerner1991,Holstein1997} formulations.

Historically, the FW transformation was developed and studied primarily in single-particle, quantum-mechanical settings. 
We therefore begin with these applications before turning to field-theoretic generalizations.
In particular, quantum-mechanical studies have focused on the explicit construction of the FW transformation. 
The original scheme~\cite{Foldy1950} progressively block-diagonalizes the Hamiltonian through a sequence of unitary transformations, systematically pushing odd operators to higher orders in the $1/m$ expansion. 
In contrast, direct methods~\cite{Pachucki2005} determine the block-diagonal Hamiltonian to a given order in a single step, without resorting to an iterative sequence of transformations. 
A third category consists of exact constructions for special classes of Hamiltonians~\cite{Eriksen1958,Eriksen1960}, for which a closed-form block-diagonalization can be obtained without relying on a perturbative $1/m$ expansion. 
Their mutual consistency, uniqueness, and relation to the FW representation~\cite{Silenko2003} have been widely discussed~\cite{Neznamov2009,Silenko2016}, in particular because different methods may yield distinct block-diagonal representations beyond a given order, while remaining related by additional unitary transformations.
This reflects the fact that there is no unique NR reduction through FW transformations in the same sense that there is no unique EFT basis, due to the presence of redundant operators.
In our work, we perform the NR reduction using iterative FW field redefinitions in a generalized scheme.
We show that this scheme reproduces the standard NRQCD/NRQED contributions at tree level and maps radiative short-distance information encoded in the relativistic form factors onto all one-gauge-field interactions up to the order $m_f^{-3}$.
Different choices of reduction schemes may lead to different parametrizations of the NR Wilson coefficients.
The closest quantum-mechanical treatment to our approach starts from an effective Dirac equation taking into account the first EFT term at the order $1/\Lambda$ and performing the standard FW transformation~\cite{Jentschura2005}.
While the spirit is similar and reproduces certain aspects of our approach, it is not a complete EFT treatment.

Next, generalizing single-particle approaches, we move to field-theoretical treatments, which often start from a Lagrangian for NR reductions of, \eg heavy-particle EFTs~\cite{Gardestig2007} such as heavy-quark effective field theory (HQET)~\cite{Isgur1989,Isgur1990,Manohar1997}.
To place HQET in the context of our approach, we briefly review it in more detail.
It is an effective field theory of QCD describing the dynamics of heavy quarks inside hadrons in the limit $m_Q \gg \Lambda_\text{QCD}$, for instance in $B$ mesons, which are bound states of a bottom antiquark and a lighter quark. 
Here, $\Lambda_\text{QCD}$ denotes the scale of non-perturbative QCD dynamics governing residual momenta and energy transfers within the hadron. 
The heavy hadron moves with four-velocity $v^\mu$, normalized as $v^2=1$, and the heavy quark momentum can be decomposed as $p_Q^\mu=m_Qv^\mu+\ell^\mu$, where $\ell^\mu \sim \Lambda_\text{QCD}$ is the residual momentum. 
In this regime, the quark is close to its mass shell, \ie $\left( p_Q^2-m_Q^2 \right)/m_Q^2=\mathcal{O}\left(\Lambda_\text{QCD} / m_Q \right)$.
One established approach~\cite{Georgi1990} to obtain the HQET Lagrangian at tree level follows from separating the heavy quark field $Q= \exp{-\ii m_Q x_\mu v^\mu} \left( H_v + h_v \right)$ into large $H_v = \exp{\ii m_Q x_\mu v^\mu}\left( \mathds{1} + \slashed{v} \right) Q/2$ and small $h_v = \exp{\ii m_Q x_\mu v^\mu} \left( \mathds{1} - \slashed{v} \right) Q/2$ components in the Lagrangian $\mathcal{L}_Q = \bar{Q} \left( \ii \slashed{D} - m_Q \right) Q$, where the phase factor removes the mass term from the large component.
Subsequently, integrating out the field $h_v$ directly or eliminating it via corresponding field redefinitions~\cite{Gardestig2007} in the path integral yields the tree-level HQET Lagrangian.\footnote{
The terminology large and small stems from the fact that, upon deriving the EOMs, one finds $h_v \sim \Lambda_\text{QCD} H_v / m_Q $.}
In a general setting of heavy-fermion EFTs, which includes also HQET, the well-known iterative FW transformation has been applied~\cite{Gardestig2007} to obtain the NR Lagrangian and compared with the result of integrating out the field $h_v$.
While the Lagrangians differ in form, these two methods are once again connected by a field redefinition~\cite{Gardestig2007}.
This discussion illustrates how the FW transformation is used in the EFT context either as an alternative route to deriving the NR tree-level Lagrangian by integrating out large components, or as a general tool applied to generic Lagrangians in the same spirit as in its original single-particle quantum-mechanical formulation.

To some extent, our approach resembles straightforward applications of the iterative FW transformation to field-theoretical Lagrangians as discussed above, but it differs in several important aspects: 

(i) First the FW transformation is often applied to only the fundamental Lagrangian, and therefore reproduces only tree-level contributions to the NR description.
In contrast, we start from a relativistic EFT, such as SMEFT~\cite{Buchmueller1986,Grzadkowski2010,Lehman2014}, HEFT~\cite{Feruglio1993}, or LEFT~\cite{Jenkins2018,Jenkins2018a,Liao2020,Murphy2021}, whose Wilson coefficients encode short-distance effects from integrated-out degrees of freedom and, provided the corresponding matching has been performed, may also contain hard contributions associated with the subsequent relativistic-to-NR scale separation.
Applying the FW reduction to such a Lagrangian allows one to obtain not only tree-level contributions but also the NR Wilson coefficients associated with short-distance information already encoded in the relativistic coefficients.
Hence, the structure of the underlying relativistic theory is directly reflected in the NR theory, and the associated constraints reminiscent of hidden Lorentz invariance emerge naturally from the reduction.
Furthermore, our generalized NR reduction scheme involves redefinitions of  gauge fields, which are essential to maintain a consistent power counting of NR operators.
Without these redefinitions, relevant contributions may be obscured or incorrectly organized in the NR expansion.
This difference is most clearly seen from the fact that NR pair-creation effects are retained within our framework. 
These effects arise already at tree level and are therefore present when applying our procedure to the standard QCD/QED Lagrangian.
While in many applications the antiparticles are eventually integrated out, they remain relevant in systems such as positronium.
To our knowledge, the combination of both aspects, namely the ability of the FW reduction to reproduce the standard NRQED/NRQCD operator structure through the orders considered, together with the consistent inclusion of NR pair-creation processes through gauge-field redefinitions, has not been discussed.
In contrast to bottom-up NR EFT constructions in which Wilson coefficients are determined by matching a selected set of amplitudes or Green functions, our approach starts from a relativistic EFT containing the QCD and QED field degrees of freedom retained in the low-energy theory and performs the NR reduction at that level.
As a consequence, contributions appear beyond those commonly discussed in the literature, such as NR one-photon-one-gluon terms.

(ii) A conceptual difference between the standard construction and our approach arises in the treatment of gauge fields.
In the conventional approach, one has the freedom to work with NR spinors that transform non-trivially under Lorentz boosts, while the gauge fields remain Lorentz covariant.
In our derivation, we redefine only the spatial components of the gauge fields.
This redefinition is not Lorentz covariant and is used to implement the EOM of electric fields.
As a consequence, the gauge field is no longer manifestly Lorentz covariant, even though a covariant redefinition is possible.
However, such a redefinition would change the parametrization of the NR Wilson coefficients and would not be directly equivalent to the standard NR Lagrangian.
In addition, it may introduce a coupling of higher orders in $m_f$ to formally lower orders.
Because of this change in parametrization, the transformation behavior of the NR spinor under Lorentz boosts also differs from the standard choice in NRQED~\cite{Heinonen2012}.
Importantly, the non-Lorentz-covariant gauge field redefinition does not pose a fundamental issue.
The redefined gauge field still acts as a gauge field under gauge transformations, and Lorentz invariance is ensured by the covariant starting point.

(iii) Regarding the matching procedure, both matching to relativistic theories such as LEFT~\cite{Jenkins2018,Jenkins2018a,Liao2020,Murphy2021} or to NR theories like NRQCD~\cite{Lepage1992,Manohar1997,Pineda1998a} and NRQED~\cite{Hill2013} are well established.
In our approach, the interplay between matching, renormalization, and the NR reduction is more subtle. The local and invertible field redefinitions used here reorganize the operator basis but do not by themselves define an EFT with a lower dynamical cutoff.
The actual transition to the NR regime occurs once external states are restricted to momenta $\abs{\vect{p}} \ll m_f$ and residual energies $\abs{p_f^0 - m_f} \ll m_f$.
This restriction is essential, since without it the $1/m_f$ expansion would not constitute a controlled NR expansion.
As a consequence, a rematching of the NR Lagrangian can only be avoided if the relativistic matching input contains the hard contributions associated with modes that are no longer dynamical in the NR EFT.
While we focus in this paper on the general implementation of the NR reduction of a relativistic EFT Lagrangian centered around the FW field redefinitions, a systematic treatment of matching and renormalization in combination with the NR reduction is left for future work.

(iv) Related to the previous point, we note that although the NR reduction in our Lagrangian can formally be applied to arbitrary fermions, the NR limit is not necessarily meaningful for every type of fermion.
For example, we may also include the lightest up and down quarks. 
However, to obtain a convergent series, we have to restrict momenta of particles to be smaller than the up-quark mass, which lies below $\Lambda_\text{QCD}$.
In contrast, the physically relevant dynamics of quarks confined in hadrons occurs at the scale of $\Lambda_\text{QCD}$, so that the resulting expansion is not applicable in this regime. 
This is also the reason why HQET requires a heavy quark and fails for light quarks.
Our particular NR EFT is thus not suited to describe up or down quarks in the NR limit and whether a meaningful NR limit exists depends on the hierarchy of scales relevant for the processes and states under consideration.
Even though we begin with a relativistic EFT Lagrangian covering all QCD and QED interactions within the considered symmetries to demonstrate the most general NR limit, the framework can equally be applied after integrating out irrelevant degrees of freedom at the relativistic level.
For instance, to obtain NRQED we can also integrate out all other particles and start with a relativistic EFT for only the electron, positron, and the photon, for which we can perform our derivation.

The central result of this work is a systematic algebraic route from a relativistic EFT to its NR counterpart through generalized Foldy-Wouthuysen-type field redefinitions. 
Applied to the QED and QCD sectors considered here, the method reproduces the established NRQED and NRQCD operator structures through the orders considered and directly relates their Wilson coefficients to those of the relativistic EFT. 
In this way, the underlying Lorentz symmetry of the relativistic theory is automatically reflected in relations among the resulting NR Wilson coefficients.
Two features of the construction are particularly relevant for its generalization beyond standard NRQED and NRQCD. 
First, the EFT is formulated at the relativistic level, where the underlying spacetime symmetries remain manifest. 
Second, the generalized reduction scheme is not restricted to QED and QCD interactions but can be applied to additional relativistic operators and fields, provided that a well-defined NR expansion exists. 
The two applications presented in section~\ref{Sec:6} illustrate these aspects. For the Poincar\'e generators, the reduction yields the first nontrivial EFT corrections to the NR boost generator and reproduces the established QED/QCD result.
In addition, we obtain the NR Wilson coefficients of a BSM axion field through the corresponding NR reduction in a simplified setting.
Once we include axion EFT terms, we find an NR axion-dependent EDM interaction, which is of the order $\left( m_f^2 c_{\partial^2 a}^{(f)} /\Lambda^2 \right) \left(m_a / m_f \right)^2$ together with an axion-induced mass shift, \ie it scales quadratically with the ratio between the axion mass and the fermion mass $m_a / m_f$ but also depends on the scaling of the EFT Wilson coefficient $\left( m_f^2 c_{\partial^2 a}^{(f)} / \Lambda^2 \right)$. 
Due to the generalized NR reduction scheme, we obtain an axion-dependent mass shift, which is in lowest order proportional to $ \left( m_f c_{\partial a}^{(f)}/\Lambda \right)^2 (m_a/m_f)^4$.
Hence, it appears also in the case of only using the lowest-order axial current for the axion model, but the ratio $m_a / m_f$ enters with the fourth power.

Finally, the scheme developed in this work is especially suited to study further BSM fields and Dirac fermions in non-inertial spacetime background fields. 
Constructing the EFT according to the same symmetries as the fundamental theory is straightforward and, thus, possible subtleties due to NR residual symmetries can be circumvented. 
Deriving the corresponding NR Lagrangians from relativistic EFT descriptions is expected to be an ideal basis for theoretical models describing NR experiments designed to test new physics like atom-interferometric setups~\cite{Dimopoulos2008Atomic_gravitational,Abe2021,Abend2023,Abdalla2025,Schaffrath2025} or quantum-clock measurements~\cite{Zych2011,Yudin2018,Loriani2019,DiPumpo2021,DiPumpo2022,Chiba2022,DiPumpo2023}.

\appendix
\section{EFT construction} \label{App:A}
In this appendix, we provide the construction of the relativistic QCED Lagrangian introduced in section~\ref{Sec:2}. 
Starting from the complete set of operators compatible with the assumed symmetries through order $\Lambda^{-3}$, we subsequently eliminate redundant operators by Lorentz-covariant field redefinitions and thereby obtain the reduced basis of eq.~\eqref{Eq:LTinv}.
\subsection{Methodology} \label{App:A1}
Constructing a relativistic EFT requires the inclusion of all operators that are allowed by the symmetries, namely Hermiticity, gauge invariance, Poincar\'e symmetries, $P$, $T$, and $C$ symmetry. 
Within the fermion-number-conserving sectors considered here, fermionic operators can be organized in terms of bilinears $\bar{f} \Gamma f$.
Gauge covariance is implemented through $D_{\mu} = \partial_\mu + \ii \gX \mathcal{A}_{\mathcal{X},\mu} $ with $\Gamma$ chosen from the Dirac basis $\mathds{1}, \gamma_5, \gamma^\mu, \gamma^\mu \gamma_5, \sigma^{\mu \nu}$, while gauge fields require the field strength tensor $\mathcal{X}_{\mu \nu}$.
We decompose the Lagrangian according to the number of fermion fields $\mathcal{L} = \sum_{f} \bar{f} O f + \sum_{f,g} \qty( \bar{f} O f ) \qty( \bar{g} O g ) + \dots + \mathcal{X}_{\mu \nu} O^{\mu \nu \alpha \beta} \mathcal{X}_{\alpha \beta} $.
The task then reduces to constructing all operators allowed by the symmetries within each sector.

A general construction rule for the operators $O$, which ensures preserved symmetries, can be summarized as follows:
\begin{itemize}
    \item Include as many factors of $iD^\mu$ as required for the corresponding term to have the correct mass dimension of an energy density at the given order in $\Lambda$
    \item Fermion fields may be combined with any of $\mathds{1},\gamma^\mu$, $\ii \gamma_5$, $\ii \gamma^\mu \gamma_5$, $\ii \sigma^{\mu \nu}$ that yields an even number of Lorentz indices
    \item Any term in the Lagrangian must be made Hermitian by adding the Hermitian conjugate
    \item Build all unique Lorentz scalars by contracting indices with the metric $\eta_{\mu \nu}$. Within the building blocks considered here,
    structures containing a single $\gamma_5$ require a pseudoscalar or pseudotensor contraction, which can be represented using a Levi-Civita tensor $\varepsilon_{\alpha \beta \mu \nu}$ 
    \item For fermion fields with color charge, any non-vanishing color trace over field strength tensors in a constructed term yields another unique term
\end{itemize}

\subsection{Full EFT Lagrangian}
Following the construction rules presented in section~\ref{App:A1}, the different sectors of the Lagrangian for QCD and QED interactions can be straightforwardly constructed.
To this end, table~\ref{tab:3} summarizes all possible combinations of gauge covariant derivatives for $\mathcal{L}_2$ that feature two operators in $\Lambda^{-1}$, three operators in $\Lambda^{-2}$, and ten in $\Lambda^{-3}$, where two of them follow from applying the color trace, not shown in the table.
\begin{table}[ht]
    \caption{All allowed symmetry-preserving operators, organized by their order in $\Lambda$ (rows) and the associated Dirac-basis matrices (columns).}
    \setlength{\tabcolsep}{0.0pt}
    \centering
    \begin{tabular}{cccccc}
    \toprule
         & scalar ($\mathds{1})$& pseudoscalar ($\gamma_5$) & vector ($\gamma^\mu$) & axial ($\gamma^\mu \gamma_5$) & tensor ($\ii \sigma^{\mu \nu}$)  \\
    \midrule 
        \rowcolor{gray!30}
    \rule{0pt}{1.5em} $\Lambda^{-1}$ & $D^2$ & & & & $ \left[ D_{\mu}, D_{\nu} \right] $ \\ [0.5em]
    $\Lambda^{-2}$ & & & \makecell[c]{ \rule{0pt}{1.5em} $ \left\{ \ii D_\mu, D^2 \right\}$ \\ [0.5em] $\ii D_\nu D_\mu D^\nu$ \\ [0.5em] } & $  \varepsilon_{\alpha \beta \nu \mu} D^\alpha D^\beta D^\nu$ & \\ [0.5em]
    \rowcolor{gray!30}
    \rule{0pt}{1.5em} $\Lambda^{-3}$ & \makecell[c]{ $D^4$ \\ [0.5em] $D_\mu D^2 D^\mu$ \\ [0.5em] $D^\mu D^\nu D_\mu D_\nu$} & $\ii \varepsilon_{\alpha \beta \mu \nu} D^\alpha D^\beta D^\mu D^\nu$ & & & \makecell[c]{ \rule{0pt}{1.5em} $ \left\{ D^2, D_\mu D_\nu \right\}$ \\ [0.5em] $ D_\mu D^2 D_\nu$ \\ [0.5em] $ D_\alpha D_\mu D_\nu D^\alpha$ \\ [0.5em] $  D_\alpha D_\mu D^\alpha D_\nu - D_\nu D^\alpha D_\mu D_\alpha $ \\ [0.5em] }\\ 
    \bottomrule
    \end{tabular}
    \label{tab:3}
\end{table}

We may cast the operators with respect to gauge-covariant derivatives depicted in table~\ref{tab:3} into a more physical form with respect to field-strength tensors $\left[ D_\mu, D_\nu \right] = \ii \gX \mathcal{X}_{\mu \nu}$. 
In this representation, the identification of redundant operator combinations becomes more subtle, and the identities
\begin{subequations}
    \begin{align}
        \left[ D^\mu, \left\{ D^\nu, \mathcal{X}_{\mu \nu} \right\} \right] &= \left\{ D^\nu, \left[ D^\mu, \mathcal{X}_{\mu \nu} \right]  \right\} + i \gY \left\{ \mathcal{Y}^{\mu \nu} , \mathcal{X}_{\mu \nu} \right\} \\
         \left\{ D^\alpha , \left\{ D_\mu , \mathcal{X}_{\nu \alpha} \right\} \right\} \sigma^{\mu \nu} &= \left\{ D_\mu, \left\{ D^\alpha, \mathcal{X}_{\nu \alpha} \right\} \right\} \sigma^{\mu \nu} + \ii \gY \left[ \mathcal{Y}_\mu^{\, \, \, \, \alpha}, \mathcal{X}_{\alpha \nu} \right] \sigma^{\mu \nu} \\
         \left[ D_\mu, \left[ D^\alpha, \mathcal{X}_{\alpha \nu} \right] \right] \sigma^{\mu \nu} &= \frac{1}{2} \left[ D_\alpha, \left[ D^\alpha, \mathcal{X}_{\mu \nu} \right] \right] \sigma^{\mu \nu} + \ii \gY \left[ \mathcal{Y}_\mu^{\, \, \, \, \alpha}, \mathcal{X}_{\alpha \nu} \right] \sigma^{\mu \nu}
    \end{align}
\end{subequations}
can be useful in identifying which operators are independent.
After these algebraic manipulations, we arrive at the two-fermion Lagrangian $\mathcal{L}_{2} = \sum_f \bar{f} \mathcal{I}_f f$ with the kernel
\begin{subequations} \label{Eq:1}
\begin{align} \label{Eq:1a}
    \mathcal{I}_f =& i \slashed{D} - m_f + \cI \frac{D^2}{\Lambda} + \cIII \frac{ \left\{ \ii \slashed{D} , D^2 \right\}}{\Lambda^2} +\cVI \frac{D^4}{\Lambda^3} + \gX \left[ -  \cII \frac{\mathcal{X}_{\mu \nu} \sigma^{\mu \nu}}{4\Lambda}  + \cIV \frac{\left[ D^\mu, \mathcal{X}_{\mu \nu} \right] }{\Lambda^2} \gamma^\nu \right] \notag \\
    &+\gX \left[ \cV \frac{\left\{\ii D^\mu , \mathcal{\tilde{X}}_{\mu \nu} \right\} }{\Lambda^2} \gamma^\nu \gamma_5 - \cXI \frac{ \left\{ D_{\mu} , \left\{ D^\alpha , \mathcal{X}_{\alpha \nu} \right\} \right\}}{2\Lambda^3} \sigma^{\mu \nu } + \cVIII \frac{ \left\{ \ii D^\nu, \left[ D^\mu, \mathcal{X}_{\mu \nu} \right] \right\}}{\Lambda^3} \right]   \notag \\
    &+ \gX \left[ \cX  \frac{ \left\{ D^2 , \mathcal{X}_{\mu \nu} \right\}}{\Lambda^3} \sigma^{\mu \nu}  + \cXII \frac{ \left[ D^\alpha, \left[ D_{\alpha} , \mathcal{X}_{\mu \nu}\right] \right]}{\Lambda^3} \sigma^{\mu \nu} \right]  +\gX \gY \ii \cXIII \frac{  {\mathcal{X}_\mu}^{\alpha} \mathcal{Y}_{\alpha \nu} \sigma^{\mu \nu} }{2\Lambda^3} \notag \\
    & + \gX \gY \left[ \mathcal{X}^{\mu \nu} \frac{ \cVII \mathcal{Y}_{\mu \nu} + \ii  \cIX \mathcal{\tilde{Y}}_{\mu \nu} \gamma_5 }{4\Lambda^3}  + \cXIV \frac{ \text{Tr} \left\{ \mathcal{X}^{\mu \nu} \mathcal{Y}_{\mu \nu} \right\}}{4\Lambda^3} + \ii  \cXV \frac{ \text{Tr} \left\{ \mathcal{X}^{\mu \nu} \mathcal{\tilde{Y}}_{\mu \nu} \right\} }{4 \Lambda^3} \gamma_5 \right] 
\end{align}
in the notation introduced in section~\ref{subsec:Notation}.
The leading-order contribution uses the Feynman-slash notation $\slashed{D} = D_\mu \gamma^\mu$ and $m_f$ is the fermion's mass.
Furthermore, we label the Wilson coefficients $c^{(f)}_{X}$ of the EFT terms by the operators $X$ which they are representing.
Similarly, the four-fermion sector involving fermionic spinor fields $(f,g)$ takes the form
\begin{align} \label{Eq:1b}
    \mathcal{L}_4 =& \frac{1}{\Lambda^2 }  \sum_{f,g}  \sum_{\Gamma} d_{\Gamma,\mathcal{X}}^{(fg)} \left( \bar{f} T^a_\mathcal{X}  \Gamma f \right) \left(\bar{g} T^a_\mathcal{X} \Gamma g \right)  + \frac{1}{\Lambda^3} \sum_{f,g} \Bigg[ d_{ 2,\mathcal{X} }^{(fg)} \qty( \bar{f} T^a_{\mathcal{X}} f ) \qty( \bar{g} T^a_{\mathcal{X}} \ii \slashed{D} g + \text{H.c.} ) \notag  \\
    &+  d_{3,\mathcal{X}}^{(fg)} \qty( \bar{f} T^a_{\mathcal{X}} \gamma_\mu f ) \qty( \bar{g} T^a_{\mathcal{X}} \ii D^\mu g + \text{H.c.}  ) + d_{4,\mathcal{X}}^{(fg)} \qty( \bar{f} T^a_{\mathcal{X}} \ii \gamma_5 f ) D_{\mu} \qty( \bar{g} T^a_{\mathcal{X}}  \gamma^\mu \gamma_5 g ) \notag \\
    &+ d_{5,\mathcal{X}}^{(fg)} \qty( \bar{f} T^a_{\mathcal{X}} \gamma_\nu f) D_\mu \qty( \bar{g} T^a_{\mathcal{X}}  \sigma^{\mu\nu} g ) + d_{6,\mathcal{X}}^{(fg)} \varepsilon_{\mu \nu \alpha \beta} \qty( \bar{f} T^a_{\mathcal{X}} \sigma^{\mu \nu} f ) \qty( \bar{g} T^a_{\mathcal{X}} \ii D^{\alpha} \gamma^\beta \gamma_5 g + \text{H.c.}) \notag \\
    & + d_{7,\mathcal{X}}^{(fg)} \varepsilon_{\mu \nu \alpha \beta} \qty( \bar{f} T^a_{\mathcal{X}} \gamma^\alpha \gamma_5 f ) \qty( \bar{g} T^a_{\mathcal{X}} \ii D^{\beta} \sigma^{\mu \nu} g + \text{H.c.}) \Bigg]
\end{align}
and contains only EFT terms that begin at the order $\Lambda^{-2}$.
As explained in section~\ref{subsec:Notation}, $\Gamma = \left\{\mathds{1},\gamma_5, \gamma^\mu, \gamma^\mu \gamma_5, \sigma^{\mu \nu} \right\}$ represents all 16 independent Dirac basis matrices.
The first term in eq.~\eqref{Eq:1b} implies the summand $\Gamma=\gamma$ of the contraction $d_{\gamma,\mathcal{X}}^{(fg)} \left( \bar{f} T^a_\mathcal{X}  \gamma_\mu f \right) \left(\bar{g} T^a_\mathcal{X} \gamma^\mu g \right)$ that yields a Lorentz scalar for instance.
Here, we again sum in addition over the photon and gluon contributions in the generator $T_\mathcal{X}^a$. 
Pure gauge field terms are summarized in the Lagrangian
\begin{align} \label{Eq:1c}
    \mathcal{L}_\text{Gauge} =&  -\frac{F^{\mu \nu} F_{\mu \nu}}{4} - \frac{\text{Tr} \left\{ \mathcal{G}^{\mu \nu} \mathcal{G}_{\mu \nu} \right\}}{2}  + c_{\text{V}F} \frac{F^{\mu \nu} \partial^2 F_{\mu\nu}}{\Lambda^2}  + c_{\text{V}\mathcal{G}}  \frac{ \text{Tr} \left\{ \mathcal{G}^{\mu \nu} D^2 \mathcal{G}_{\mu\nu} \right\}  }{\Lambda^2} \notag \\
    &+\ii g_s c_{\mathcal{G}^3} \frac{ \text{Tr} \left\{ {\mathcal{G}^{\mu}}_{\nu} {\mathcal{G}^{\nu}}_{\alpha} {\mathcal{G}^\alpha}_{\mu} \right\} }{\Lambda^2},
\end{align}
\end{subequations}
where $F_{\mu \nu}$ is the electromagnetic field strength tensor without the generator, \ie the standard $F_{\mu \nu} = \partial_\mu A_\nu - \partial_\nu A_\mu$.\footnote{The last pure gauge-field term in eq.~\eqref{Eq:1c} contains only the gluon because it is algebraically zero for the photonic field strength tensor $F_{\mu\nu}$.}

In principle, eq.~\eqref{Eq:1} can already be used as a starting point to derive the NR limit.
However, this basis of operators still contains some degree of redundancy that we will eliminate in the following to simplify calculations.

\subsection{Lorentz-covariant field redefinitions}
To reduce the number of operators in eq.~\eqref{Eq:1}, we perform local and invertible field redefinitions that leave the $S$-matrix invariant~\cite{Kamefuchi1961,Arzt1995}.
Moreover, we restrict ourselves to symmetry-preserving field redefinitions such that the redefined fields remain Lorentz covariant and behave the same under discrete symmetry operations.
The Lagrangian itself is constructed from all possible Lorentz scalars and symmetry-preserving operators and can be expressed as $\mathcal{L} \propto \bar{f} O f$ for the two- and four-fermion sector, where the latter implies that $O$ has to contain additional fermion fields.
Hence, we can use any operator $O$ from the Lagrangian in a given order to perform a redefinition of the fermion field and to eliminate terms in the next order via the (non-unitary) transformation $f = \left( 1 +O/ \Lambda \right) f^\prime$.
While these redefinitions can also be expressed as an exponential expanded to the relevant order, for the elimination of redundant terms a linear field redefinition suffices.
For gauge fields, we may choose any vector that transforms covariantly under Lorentz transformations while preserving the gauge transformation behavior of the field.
A symmetry-preserving set of linear field redefinitions sufficient to eliminate the redundant operators considered here is given by
\begin{subequations} \label{Eq:fieldredefs}
    \begin{align} 
        f =& \Bigg( 1 + \frac{a_{D}^{(f)} \ii \slashed{D}}{\Lambda} +  \frac{a_{D^2}^{(f)} D^2 + a_\mathcal{X}^{(f)} \gX \mathcal{X}_{\mu \nu} \sigma^{\mu \nu}}{\Lambda^2} + \frac{a_{D^3}^{(f)} \left\{ \ii \slashed{D}, D^2 \right\} + a_{D\mathcal{X}}^{(f)} \gX \left[ D^\mu , \mathcal{X}_{\mu \nu} \right] \gamma^\nu }{\Lambda^3} \notag\\
        &+\frac{a_{D\mathcal{\tilde{X}}}^{(f)} \gX \left\{ \ii D^\mu , \mathcal{\tilde{X}}_{\mu \nu} \right\} \gamma^\nu \gamma_5 }{\Lambda^3} + \frac{1}{\Lambda^3} \sum_{g, \Gamma} a_{\Gamma,\mathcal{X}}^{(fg)} T^a_{\mathcal{X}} \Gamma (\bar{g}^\prime T^a_{\mathcal{X}} \Gamma g^\prime)  \Bigg) f^\prime \\
        A_{\mathcal{F},\mu} =& A^{\prime}_{\mathcal{F},\mu} + e \sum_{f} Z_f \left[ b_{\gamma,\mathcal{F}}^{(f)} \frac{ \bar{f} \gamma_\mu f}{\Lambda^2} + b_{D,\mathcal{F}}^{(f)}  \frac{ \bar{f} \ii D_\mu f + \text{H.c.} }{\Lambda^3}  + b_{\sigma,\mathcal{F}}^{(f)}  \frac{\partial^\nu \left( \bar{f} \sigma_{\nu \mu} f \right)}{\Lambda^3} \right] + b_\mathcal{F} \frac{ \partial^\nu F_{\nu \mu} }{\Lambda^2}  \\
        \mathcal{A}_{\mathcal{G},\mu} =& \mathcal{A}^{\prime}_{\mathcal{G},\mu} + g_s T^a \sum_{f} \left[ b_{\gamma,\mathcal{G}}^{(f)} \frac{ \bar{f} \gamma_\mu T^a f}{\Lambda^2} + b_{D,\mathcal{G}}^{(f)}  \frac{ \bar{f} T^a \ii D_\mu f + \text{H.c.}}{\Lambda^3}  +  b_{\sigma,\mathcal{G}} ^{(f)}  \frac{ D^\nu \left( \bar{f} T^a  \sigma_{\nu \mu} f \right)}{\Lambda^3} \right] \notag \\
        &+ b_\mathcal{G} \frac{ D^\nu \mathcal{G}_{\nu \mu}}{\Lambda^2}.
    \end{align}
\end{subequations}
These field redefinitions manipulate the Lagrangian through redefinition coefficients $a$ and $b$, which are real parameters.\footnote{We suppress the superscript $a$ representing internal components for the electromagnetic four-vector potential $A^a_{\mathcal{F}, \mu} = A_{\mathcal{F},\mu}$ for clarity.}
Once we insert the field redefinitions from eq.~\eqref{Eq:fieldredefs} into the Lagrangian, eq.~\eqref{Eq:1}, the Wilson coefficients $(c,d)$ of all operators will be modified by contributions depending on the field-redefinition parameters $(a,b)$.
Through a suitable choice of these field redefinitions, we may eliminate certain operators in the Lagrangian.
For the field redefinitions of the fermion fields, it suffices to calculate the contribution from the leading-order terms $\left\{ O/\Lambda, \ii \slashed{D} - m_f \right\}$, because higher-order contributions only make the choice of field redefinition parameters more involved but do not change the number of independent operator coefficients that can be eliminated.
However, due to the leading-order fermionic mass term $\bar{f} m_f f$ in the Lagrangian, their higher orders are coupled to lower orders, \ie field redefinitions of the form $O / \Lambda^j$ may imply a contribution of the form $O m_f/ \Lambda^j$.
As a result, the equations that the field redefinition parameters $(a,b)$ must satisfy to remove operators from the Lagrangian become coupled.
To solve these equations, one has to impose further constraints on the $(c,d)$ Wilson coefficients appearing in the Lagrangian.
For perturbatively suppressed Wilson coefficients, the constraint equations remain continuously connected to their leading-order solution and are therefore nonsingular within the perturbative regime considered here.
Even if some constraint equations were to become singular in a non-perturbative setting, the corresponding operators can simply be retained, together with their Wilson coefficients, resulting in a larger reduced basis.
Therefore, we will not examine these particular constraints in more detail.
Consequently, what matters is which terms $(a,b)$ from the field redefinition are connected to the terms $(c,d)$ in the Lagrangian.
Table~\ref{tab:1} summarizes this mapping: on the left, we list the operators in the Lagrangian labeled by the $(c,d)$ Wilson coefficients, and on the right, we list all field-redefinition operators whose parameters $(a,b)$ also enter the corresponding Lagrangian operator.

\begin{table}[ht]
    \caption{The operator columns show the operators in the Lagrangian labeled by the $(c,d)$ Wilson coefficients and the $(a,b)$ columns summarize which field redefinition parameters $(a,b)$ enter these operators. Underlined Wilson coefficients $\underline{c}$ and $\underline{d}$ imply that the corresponding operators are eliminated through a suitable choice of the respective underlined redefinition parameters $\underline{a}$ or $\underline{b}$.}
    \label{tab:1}
    \setlength{\tabcolsep}{0.0pt}
    \setlength{\extrarowheight}{6pt}
    \centering
    \begin{tabular}{ >{\hspace{4pt}} c <{\hspace{4pt}} >{\hspace{4pt}} c <{\hspace{4pt}} || >{\hspace{4pt}} c <{\hspace{4pt}}  >{\hspace{4pt}} c <{\hspace{4pt}}  || >{\hspace{4pt}} c <{\hspace{4pt}}  >{\hspace{4pt}} c <{\hspace{4pt}}  }
    \toprule
      Operator & $(a,b)$ & Operator & $(a,b)$ & Operator & $(a,b)$   \\
    \midrule
    \rowcolor{gray!30}
        \rule{0pt}{1.8em} $c_{\slashed{D}}$ & $a_D^{(f)}$ & $\underline{\cVIII}$ & $ a_{D^3}^{(f)}, a_{D \mathcal{A}}^{(f)}, \underline{b_{D,\mathcal{A}}^{(f)}}$ & $c_{G^3}$ & $b_G$ \\ [0.5em]
         \rule{0pt}{1.8em} $\underline{\cI}$ & $\underline{a_D^{(f)}}$ &  $\cIX$ & $ a_{D \mathcal{\tilde{A}}}^{(f)} $ & $\underline{d_{2,\mathcal{X}}^{(fg)}}$ & $\underline{a_{1,\mathcal{X}}^{(fg)}}$ \\ [0.5em]
    \rowcolor{gray!30}
        \rule{0pt}{1.8em} $\cII$  & $a_D^{(f)}$ & $\underline{\cX}$ & $a_{D^3}^{(f)}, \underline{a_{D \mathcal{\tilde{A}}}^{(f)}}$ & $\underline{d_{3,\mathcal{X}}^{(fg)}}$ & $b_{D,\mathcal{X}}^{(f)}, \underline{a_{\gamma, \mathcal{X}}^{(fg)}}$  \\ [0.5em]
        $\underline{\cIII}$ & $\underline{a_{D^2}^{(f)}}$ & $\cXI$ & $ a_{D^3}^{(f)}, a_{D\mathcal{\tilde{A}}}^{(f)} $ & $\underline{d_{4,\mathcal{X}}^{(fg)}}$ & $\underline{a_{\gamma_5, \mathcal{X}}^{(fg)}}, a_{\gamma \gamma_5,\mathcal{X}}^{(fg)}$ \\ [0.5em]
    \rowcolor{gray!30}
       \rule{0pt}{1.8em} $\cIV$ & $ a_\mathcal{A}^{(f)}, b_{\mathcal{A}}, b_{ \gamma, \mathcal{A}}^{(f)}$ &  $\underline{\cXII}$ & $ \underline{a_{D \mathcal{A}}^{(f)}}, a_{D \mathcal{\tilde{A}}}^{(f)}, b_{\sigma,\mathcal{A}}^{(f)}$ & $\underline{d_{5,\mathcal{X}}^{(fg)}}$ & $\underline{b_{\sigma,\mathcal{X}}^{(f)}}, a_{\sigma, \mathcal{X}}^{(fg)},a_{\gamma, \mathcal{X}}^{(fg)}$  \\ [0.5em]
        \rule{0pt}{1.8em} $\underline{\cV}$ & $\underline{a_\mathcal{A}^{(f)}}$ & $\cXIII$ & $a_{D \mathcal{A}}^{(f)}, a_{D\mathcal{\tilde{A}}}^{(f)}, b_{\sigma, \mathcal{A}}^{(f)}$ & $\underline{d_{6,\mathcal{X}}^{(fg)}}$ & $\underline{a_{\sigma, \mathcal{X}}^{(fg)}}$ \\ [0.5em]
    \rowcolor{gray!30}
    \rule{0pt}{1.8em} $\underline{\cVI}$ & $\underline{a_{D^3}^{(f)}}$ & $\underline{c_{\text{V}\mathcal{X}}}$ & $\underline{b_{\mathcal{X}}}$ & $\underline{d_{7,\mathcal{X}}^{(fg)}}$ & $\underline{a_{\gamma \gamma_5,\mathcal{X}}^{(fg)}}$ \\ [0.5em]
    \rule{0pt}{1.8em} $\cVII$ & $a_{D^3}^{(f)}$ & $\underline{d_{\gamma,\mathcal{X}}^{(ff)}}$ & $\underline{b_{\gamma,\mathcal{X}}^{(f)}}$ && \\ [0.5em]
    \bottomrule
    \end{tabular}
\end{table}
Because several operators in the Lagrangian depend on multiple field-redefinition parameters, there is not a unique reduced basis.
Our choice of field redefinitions is highlighted by underlined Wilson coefficients and field redefinition parameters.
For instance, the operator proportional to the Wilson coefficient $\cI$ is eliminated ($\underline{\cI}$ underlined) by a suitable choice of the field redefinition parameter $a_D^{(f)}$ ($\underline{a_D^{(f)}}$ underlined), while we retain in our basis the EFT term proportional to $\cII$ (not underlined).
Formally removing $\cI$ with a field redefinition parameter $a_D^{(f)}$ introduces a modification of $c_{\slashed{D}} \neq 1$, \ie of the leading-order Lagrangian $\bar{f} \ii \slashed{D} f$.
To restore the normalization of the leading-order Lagrangian to unity, we can always rescale fermion fields in the Lagrangian by a constant.
Using this procedure, we can absorb the effects of this rescaling into redefined Wilson coefficients, with the resulting mass parameter identified as the physical mass.
With this choice of field redefinition parameters, eq.~\eqref{Eq:LTinv} follows as the reduced EFT basis.

\acknowledgments
We are grateful to W. P. Schleich for his continuing support.
We also thank A. Friedrich and G. Paz, as well as the QUANTUS team for fruitful and interesting discussions.
The authors also acknowledge contributions in the form of discussions from the Terrestrial Very-Long-Baseline Atom Interferometry (TVLBAI) proto-collaboration.
This research is supported by the Japan Society for the Promotion of Science (JSPS) (Grant Nos. JP25KJ1313, JP24K21526, JP25K00012, JP25K01691, JP25K01694, JP26K01332, JPJSJRP20221202) and the Research Foundation for Opto-Science and Technology.
The QUANTUS project is supported by the German Space Agency at the German Aerospace Center (Deutsche Raumfahrtagentur im Deutschen Zentrum f\"ur Luft- und Raumfahrt, DLR) with funds provided by the Federal Ministry for Economic Affairs and Climate Action (Bundesministerium f\"ur Wirtschaft und Klimaschutz, BMWK) due to an enactment of the German Bundestag under Grant No. 50WM2450D and No. 50WM2450E (QUANTUS-VI).

\section*{Author declarations}
\subsection*{Conflict of interest}

\noindent The authors have no conflicts to disclose.

\subsection*{Author contributions}
 \noindent{\sffamily\small\textbf{Tobias Asano}} Conceptualization (equal); Methodology (leading); Formal analysis (leading); Validation (equal); Writing - original draft (leading); Writing – review and editing (leading); Visualization (leading).
 {\sffamily\small\textbf{Fabio Di Pumpo}} Conceptualization (equal); Methodology (supporting); Validation (equal); Formal analysis (supporting); Writing - original draft (supporting); Writing – review and editing (supporting); Visualization (supporting); Supervision (leading).
 {\sffamily\small\textbf{Enno Giese}} Conceptualization (equal); Methodology (supporting); Validation (equal); Formal analysis (supporting); Writing - original draft (supporting); Writing – review and editing (supporting); Visualization (supporting); Supervision (leading).
 {\sffamily\small\textbf{Motoaki Bamba}} Conceptualization (supporting); Methodology (supporting); Formal analysis (supporting); Validation (supporting); Writing - original draft (supporting); Writing – review and editing (supporting); Visualization (supporting); Supervision (supporting).

\section*{Data availability}
\noindent The data that support the findings of this study are available within the article.

\bibliographystyle{JHEP}
\bibliography{biblio.bib}

\end{document}